\documentclass[aps,prl,reprint,showpacs,superscriptaddress,longbibliography]{revtex4-2}
\usepackage{mathtools,amssymb,graphicx,units}

\usepackage[plainpages=false,pdfpagelabels,colorlinks=true,linkcolor=red,urlcolor=blue,citecolor=blue,pdftitle={Title},pdfauthor={},pdfdisplaydoctitle=true,pdfduplex=DuplexFlipLongEdge]{hyperref}
\usepackage{verbatim}
\usepackage[english]{babel}
\usepackage[dvipsnames]{xcolor}
\usepackage[normalem]{ulem}

\graphicspath{{./}}
\nocite{*}
\definecolor{darkred}{rgb}{0.80,0,0}
\definecolor{blood}{rgb}{0.50,0,0}
\definecolor{brightred}{rgb}{1,0,0}
\definecolor{orange}{rgb}{1,0.3,0}
\definecolor{bluegreen}{rgb}{0,0.5,0.5}
\definecolor{lightblue}{rgb}{0,0.5,0.8}
\definecolor{darkgreen}{rgb}{0,0.5,0}
\definecolor{green}{rgb}{0,0.70,0}
\definecolor{darkblue}{rgb}{0,0,0.80}
\definecolor{magenta}{rgb}{1,0,1}
\definecolor{softmagenta}{rgb}{0.85,0.1,0.6}
\definecolor{mauve}{rgb}{0.6,0.1,1}
\definecolor{white}{rgb}{1,1,1}
\definecolor{black}{rgb}{0,0,0}

\newcommand{\prlsection}[1]{\noindent\textit{#1.---}\hspace{0.25em}}

\begin{document}

\title{Sign-Resolved Statistics and the Origin of Bias in Quantum Monte Carlo}
\author{Ryan Larson}
\affiliation{Department of Physics, University of California,
Davis, CA 95616, USA}
\author{Rubem Mondaini}
\email{rmondaini@uh.edu}
\affiliation{Department of Physics, University of Houston, Houston, Texas 77004, USA}
\affiliation{Texas Center for Superconductivity, University of Houston, Houston, Texas 77204, USA}
\author{Richard T. Scalettar}
\email{scalettar@physics.ucdavis.edu}
\affiliation{Department of Physics, University of California,
Davis, CA 95616, USA}

\begin{abstract}
Quantum simulations are a powerful tool for exploring strongly correlated many-body phenomena. Yet, their reach is limited by the fermion sign problem, which causes configuration weights to become negative, compromising statistical sampling. In auxiliary-field Quantum Monte Carlo  calculations of the doped Hubbard model, neglecting the sign ${\cal S}$ of the weight leads to qualitatively wrong results --- most notably, an apparent suppression rather than enhancement of $d$-wave pairing at low temperature. Here we approach the problem from a different perspective: instead of identifying negative-weight paths, we examine the statistics of measured observables in a sign-resolved manner. By analyzing histograms of key quantities (kinetic energy, antiferromagnetic structure factor, and pair susceptibilities) for configurations with ${\cal S}=\pm1$, we derive an exact relation linking the bias from ignoring the sign to the difference between sign-resolved means, $\Delta\mu$, and the average sign, $\langle {\cal S}\rangle$. Our framework provides a precise diagnostic of the origin of measurement bias in Quantum Monte Carlo and clarifies why observables such as the $d$-wave susceptibility are especially sensitive to the sign problem.
\end{abstract}

\maketitle

\prlsection{Introduction}
The fermion sign problem (SP) remains the central obstacle to extending quantum Monte Carlo (QMC) simulations of correlated electrons across many disciplines to low temperatures and large lattices\,\cite{hammond94,ceperley95,foulkes01,degrand06,carlson15,needs20}. When configuration weights acquire negative values, importance sampling becomes exponentially inefficient\,\cite{troyer05,loh90}. Considerable effort has therefore gone into its characterization\,\cite{iglovikov15}, into identifying `sign-free' schemes by virtue of special symmetries\,\cite{loh90,wu05,li16,chandrasekharan10,berg12,wang15,li19,Wang2014,ZiXiang2015}, mitigating algorithms\,\cite{Hangleiter2020,Wan2020}, or even using the SP as a tool to capture critical behavior\,\cite{wessel2017efficient,mondaini22,mou22,mondaini23,tiancheng24}. Yet a fundamental conceptual question persists: what precisely distinguishes configurations of opposite sign, and how does that distinction bias measured observables?

In world-line QMC\,\cite{hirsch82}, the answer is transparent: The sign counts the number of fermionic exchanges undergone as the world-lines propagate in imaginary time. Determining which configurations have negative weights in auxiliary field quantum Monte Carlo (AFQMC) is considerably more challenging, although possible with certain specialized Hubbard-Stratonovich (HS) transformations\,\cite{batrouni93}. In general, however, the rapid variation of the HS field in space and imaginary time makes discerning structures that lead to specific values of the sign difficult~\footnote{In models of interacting electrons and phonons, such as the Holstein Hamiltonian, the phonon kinetic energy $\hat p_{i}^{2}$ controls the imaginary time fluctuations of the phonon field, leading to a substantial reduction in the presence of configurations for which the fermion determinant is negative.}.

Here, we adopt a different approach: Rather than classifying configurations, we ask whether there is a linkage between the values of physical observables and the sign of the configuration from which the measurements arise. We do so in the specific context of Determinant Quantum Monte Carlo (DQMC)\,\cite{blankenbecler81} for the fermionic Hubbard model on a square lattice, arguably the canonical starting point to understand unconventional superconductivity\,\cite{scalapino94, Scalapino2012}. We also demonstrate that if the histograms of measurements were identical in the sign ${\cal S}=+1$ and ${\cal S}=-1$ sectors, ignoring the SP would lead to {\it correct} values of physical observables. Since this is demonstrably not the case, especially for the $d$-wave pairing susceptibility\,\cite{white89a}, the histograms must, necessarily, differ. From this analysis, we derive an exact relation connecting the bias incurred when the sign is ignored to the difference of sign resolved means, $\Delta\mu$, and the average sign $\langle S\rangle$ [see Eq.~\eqref{eq:sign_no_sign_relation}]. Whereas the histograms for ${\cal S}=+1$ and ${\cal S}=-1$ asymptotically converge at low temperatures, the simultaneous vanishing of $\langle {\cal S}\rangle$ amplifies their small differences, causing `sign-ignorant' sampling to yield increasingly incorrect observables. This approach provides a new, measurable, and model-independent diagnostic of the SP bias that applies to all reweighted QMC schemes, not just DQMC.

In what follows, we apply this framework to the Hubbard model, clarifying which observables are most sensitive to the SP and why ignoring the sign yields qualitatively incorrect results for $d$-wave pairing. 

\prlsection{Model and Methodology}
The Hubbard Hamiltonian,
\begin{align}
\hat {\cal H} = 
&- t\sum_{\langle ij\rangle, \,\sigma}  
\big( \, \hat c^{\dagger}_{i\sigma} \hat c^{\phantom{\dagger}}_{j\sigma}
+\hat c^{\dagger}_{j\sigma} \hat c^{\phantom{\dagger}}_{i\sigma} \, \big)
- \mu \sum_{i \sigma} \, \hat n_{i\sigma} 
\nonumber \\
&+ U \sum_{i} \left(\hat n_{i\uparrow} -\frac{1}{2} \right) 
\, \left(\hat n_{i\downarrow} - \frac{1}{2}\right)\ ,
\label{eq:ham_spinful}
\end{align}
is a widely studied model of metal-insulator transitions, magnetism, and exotic superconductivity\,\cite{montorsi1992hubbard,tasaki1998hubbard,fazekas1999lecture,arovas2022hubbard}. In Eq.\,\eqref{eq:ham_spinful}, $\hat c^{\dagger}_{j\sigma} (\hat c^{\phantom{\dagger}}_{j\sigma} )$ are creation (annihilation) operators at site $j$ with spin $\sigma$, $\hat n^{\phantom{\dagger}}_{j\sigma} = \hat c^{\dagger}_{j\sigma} \hat c^{\phantom{\dagger}}_{j\sigma}$ is the fermionic number operator, and we have written the interaction term in particle-hole symmetric form so that the chemical potential $\mu=0$ corresponds to half-filling ($\langle \hat n_{i\sigma} \rangle = 1/2$) for arbitrary interaction strengths $U$ and temperature $T$ on bipartite lattices. We set $t$, the hopping integral between nearest-neighbor sites, as our energy scale, and study square lattices with $L^2$ sites.

In the DQMC method, the partition function ${\cal Z} = {\rm Tr} \, e^{-\beta \hat {\cal H}}$ is expressed as a path integral by discretizing the imaginary time $\beta = L_\tau \, \Delta \tau$, and one introduces a Hubbard-Stratonovich (HS) field to decouple the interaction on each space-time site, so that $\cal Z$ becomes a sum over HS configurations ${\cal C}$ with weight ${\cal W}({\cal C})=\prod_{\sigma=\uparrow,\downarrow}\det {\cal M}_\sigma({\cal C})$, where ${\cal M}_\sigma=1+B_{L_\tau,\sigma}\cdots B_{1,\sigma}$ and $B_{\ell,\sigma}$ are single-time-slice propagators. Observables are measured from the accumulations of combinations of equal-time Green's functions $[{\cal G}_{\sigma}]_{ij} = \langle \hat c^{\phantom{\dagger}}_{i \sigma} \hat c^{\dagger}_{j \sigma} \rangle
=[{\cal M}_{\sigma}^{-1}]_{ij}$~\cite{blankenbecler81, Hirsch1985}.

For a generic, configuration-dependent observable ${\cal O}({\cal C})$, the physical expectation is
\begin{align}
    \langle{\cal O}\rangle_{\cal W}
    =\frac{\sum_{\cal C}{\cal W}({\cal C})\,{\cal O}({\cal C})}{\sum_{\cal C}{\cal W}({\cal C})}\ .
    \label{eq:expect1}
\end{align}
When ${\cal W}$ can be negative, it is convenient to write ${\cal S}({\cal C})\equiv {\cal W}({\cal C})/|{\cal W}({\cal C})|=\pm1$ and reweight by $|{\cal W}|$~\cite{Hirsch1985},
\begin{align}
    \langle{\cal O}\rangle_{\cal W}
    =\frac{\langle {\cal S}\,{\cal O}\rangle_{|{\cal W}|}}{\langle {\cal S}\rangle_{|{\cal W}|}},
    \label{eq:expect2}
\end{align}
with averages taken under the non-negative measure $|{\cal W}|$, which can be properly done with Monte Carlo sampling. Unfortunately, this rewriting in Eq.\,\eqref{eq:expect2} does not prove completely useful because the expectation values both become (exponentially) small as the inverse temperature $\beta$ and the spatial size are increased\,\cite{loh90,troyer05,iglovikov15}. Since statistical error bars remain present, they exceed the measured values, resulting in a decline in the signal-to-noise ratio as $\beta$ and the volume increase. In practice, in a Monte Carlo time series $\{{\cal C}_t\}_{t=1}^M$ sampled from $|{\cal W}|$,
\begin{align}
    \langle{\cal O}\rangle_{\cal W}
    =\frac{\frac{1}{M}\sum_t {\cal S}_t{\cal O}_t}{\frac{1}{M}\sum_t {\cal S}_t}
    =\frac{\sum_{t,+}{\cal O}_t-\sum_{t,-}{\cal O}_t}{N_+-N_-},
    \label{eq:expect3}
\end{align}
where ${\cal S}_t={\cal S}({\cal C}_t)$, ${\cal O}_t={\cal O}({\cal C}_t)$, $N_\pm$ is the number of configurations in the sampling with ${\cal S}_t=\pm1$, and $\sum_{t,\pm}$ restricts the sum to $\pm$-sign sectors.

Equation~\eqref{eq:expect3} motivates our central diagnostic. Let $P_\pm ({\cal O})$ denote the empirical distributions of ${\cal O}$ measured on configurations with ${\cal S}=\pm1$ under the $|{\cal W}|$ sampling. If $P_+=P_-$, i.e., the distributions of the physical quantities are identical, then in the large-sample limit $\sum_{t,+} {\cal O}(C_{t}) = \frac{N_{+}}{M} \sum_{t} {\cal O}({\cal C}_{t})$ and $\sum_{t,-} {\cal O}(C_{t}) = \frac{N_{-}}{M} \sum_{t} {\cal O}({\cal C}_{t})$ where $\sum_{t} {\cal O}({\cal C}_{t})$ is a sum over all configurations, regardless of sign. Inserting these expressions into Eq.~\eqref{eq:expect3} we conclude that $\langle {\cal O} \rangle_{{\cal W}} = \frac{1}{M} \sum_{t} {\cal O}({\cal C}_{t})$, precisely what one would obtain from ignoring the sign. Conversely, if $P_+\neq P_-$, the `sign-ignorant' estimate $\langle{\cal O}\rangle_{|{\cal W}|}$ is biased, i.e., $\langle{\cal O}\rangle_{{\cal W}} \neq \langle{\cal O}\rangle_{|{\cal W}|}$. In summary, this establishes the {\it histogram equivalence criterion}: Under $|{\cal W}|$ sampling, $\langle{\cal O}\rangle_{\cal W}=\langle{\cal O}\rangle_{|{\cal W}|}$ if and only if the sign-resolved distributions $P_+$ and $P_-$ of ${\cal O}$ coincide.

This criterion underlies our analysis, and in what follows, we quantify the difference between $P_\pm$ by comparing their means and via distributional distances, further connecting these differences to the bias incurred by neglecting the sign.
\begin{figure}[t]
\includegraphics[width=0.99\columnwidth]{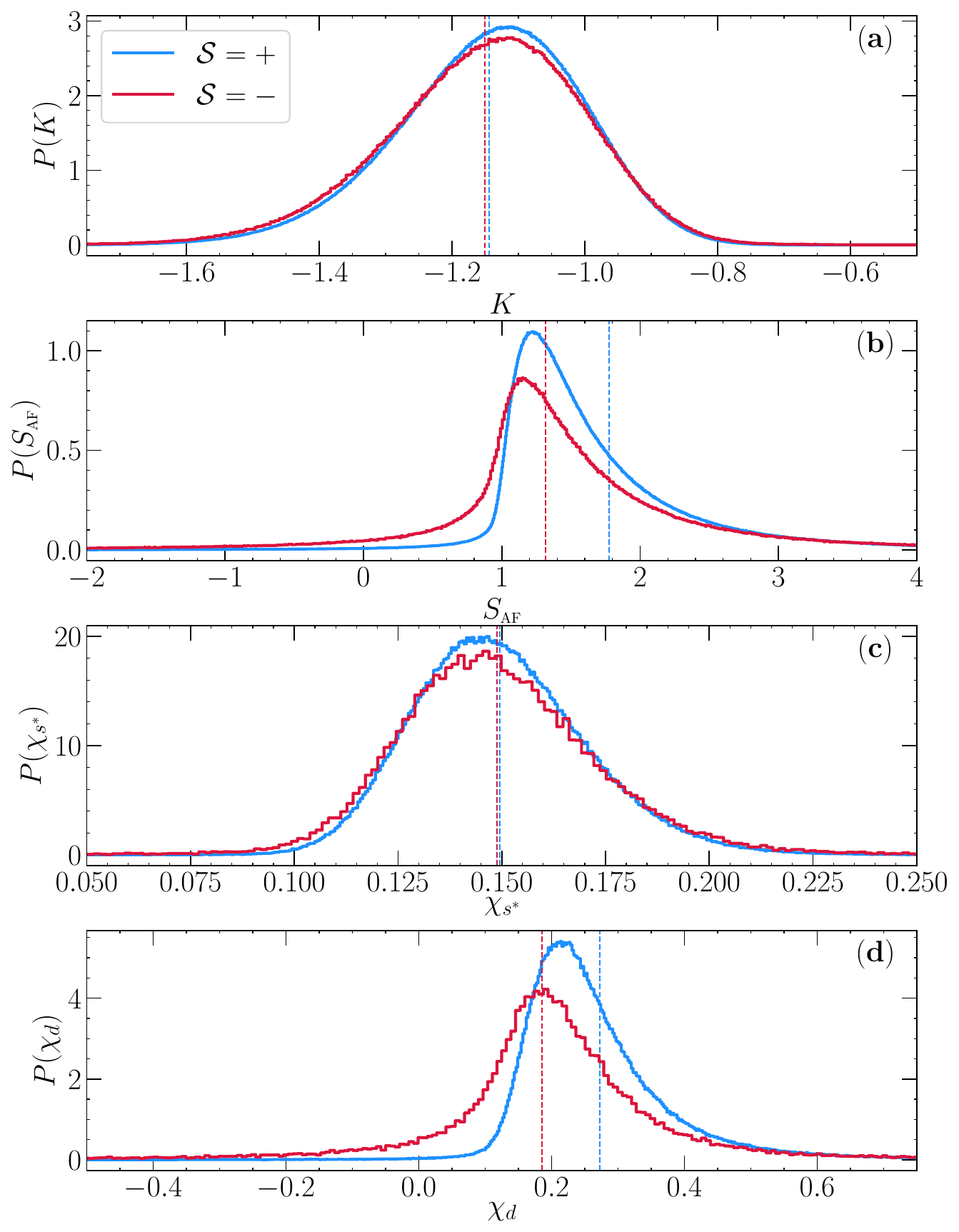}
\caption{Probability density distributions of physical observables: (a) the kinetic energy $K$, (b) the antiferromagnetic structure factor $S_{\rm \scriptscriptstyle AF}$, and the $s^*$- and $d$-wave and pair susceptibilities, $\chi_{s^*}$ and $\chi_d$, in (c) and (d). These are resolved by the value of the sign of the corresponding weight, and vertical lines give the mean values, $\mu_\pm$. All data are computed for an $8\times 8$ spatial lattice with $U/t = 6$, $T/t=1/3$ and $\mu/t=-1.4$; this leads to a mean density  $\langle \hat n\rangle \simeq 0.88$ with $\langle {\cal S} \rangle \simeq 0.83$. The imaginary-time discretization is set at $t\Delta\tau = 0.05$.}
\label{fig:fig_1_mod}
\end{figure}

\prlsection{Histograms and their dissimilarities}
Figure \ref{fig:fig_1_mod} shows the sign-resolved probability densities of representative observables: kinetic energy $K$, antiferromagnetic structure factor $S_{\rm \scriptscriptstyle AF}$, and $s^*$- and $d$-wave pair susceptibilities, $\chi_{s^*}$ and $\chi_d$, at temperature $T/t=1/3$, intermediate coupling $U/t=6$, and density $\rho= (1/L^2)\sum_{i,\sigma}\langle \hat n_{i\sigma}\rangle = 0.88$, values of relevance to modeling cuprate materials~\footnote{In more refined treatments relevant to the physics of the cuprates, a next-near-neighbor hopping $t^{\prime}\simeq-0.2 t$ is often included in the hole-doped regime~\cite{huang2018stripe, Xu2024}.}. For these parameters, the average sign is $\langle {\cal S} \rangle \simeq 0.83$. The histograms for the two signs are noticeably different, with the general trend that for the two most relevant global observables for the repulsive Hubbard model, $S_{\rm \scriptscriptstyle AF}$ and $\chi_d$, the ${\cal S}=-1$ distributions are shifted to significantly smaller values, as is emphasized by the means, shown as dashed lines. Histograms for other observables are provided in the Supplemental Materials (SM)\,\cite{SM}, along with data for other lattice sizes and Trotter discretizations.

\begin{figure}[t]
\includegraphics[width=1\columnwidth]{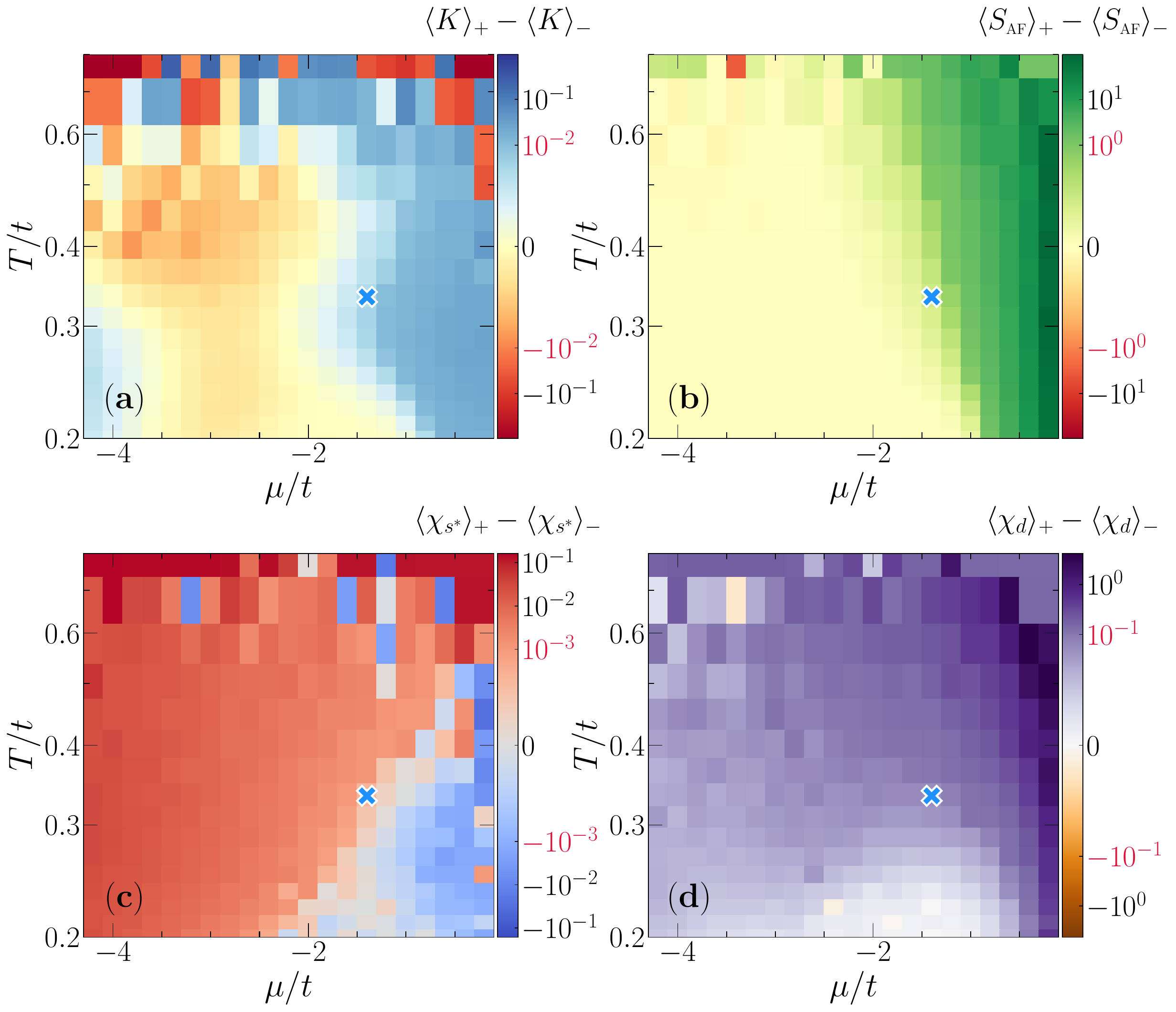}
\caption{Difference of the means $\Delta \mu_{\cal O} = \langle {\cal O}\rangle_+ - \langle {\cal O}\rangle_-$ in the $T$ vs.~$\mu$ plane. The observables $\cal O$ are the same as in Fig.~\ref{fig:fig_1_mod}: (a) the kinetic energy $K$, (b) the antiferromagnetic structure factor $S_{\rm \scriptscriptstyle AF}$, (c) the $s^*$-wave pair susceptibility $\chi_{s^*}$ and (d) the $d$-wave pair susceptibility $\chi_d$. The colored marker indicates the set of parameters chosen for Fig.~\ref{fig:fig_1_mod}, and the colors are derived from a symmetric log scale centered at zero, for enhanced visualization, with a linear range extending to the red-colored tick label in the color bar. Where applicable, parameters are as in Fig.~\ref{fig:fig_1_mod}.}
\label{fig:fig_2_mod}
\end{figure}

%% Notice, however, that
The key diagnostic of bias in the measured observable resides in the difference between the mean values $\mu_\pm=\frac{1}{N_\pm}\sum_{t,\pm}{\cal O}_t = \langle {\cal O}\rangle_\pm$. This can be seen from Eq.~\eqref{eq:expect3}, which can be simplified to $\langle {\cal O}\rangle_{\cal W} = \frac{N_+ \mu_+ - N_-\mu_-}{N_+ - N_-}$, showing that the expectation value depends only on these means and the relative populations $N_\pm$. Hence, if $\mu_+ = \mu_-$, the sign problem would not bias the expectation value, even if the detailed shapes of the histograms differed. To get a broader picture, we show in Fig.~\ref{fig:fig_2_mod} the difference in means $\Delta \mu = \mu_+ - \mu_-$ in the $T$--$\mu$ plane at fixed $U=6t$. An important general trend is that $\Delta \mu$ is typically larger in magnitude at high-$T$, but significant changes occur at different chemical potentials.

Nevertheless, distinct histogram shapes often imply different higher moments, which can influence fluctuations and correlations between observables, providing additional insight into how the sign problem manifests at the level of distributions. In Fig.~\ref{fig:fig_1_mod}, both the shift of the ${\cal S}=-1$ histograms to lower values and their broader widths signal that the negative-weight configurations sample a statistically distinct sector of configuration space.

\begin{figure}[t]
\includegraphics[width=1\columnwidth]{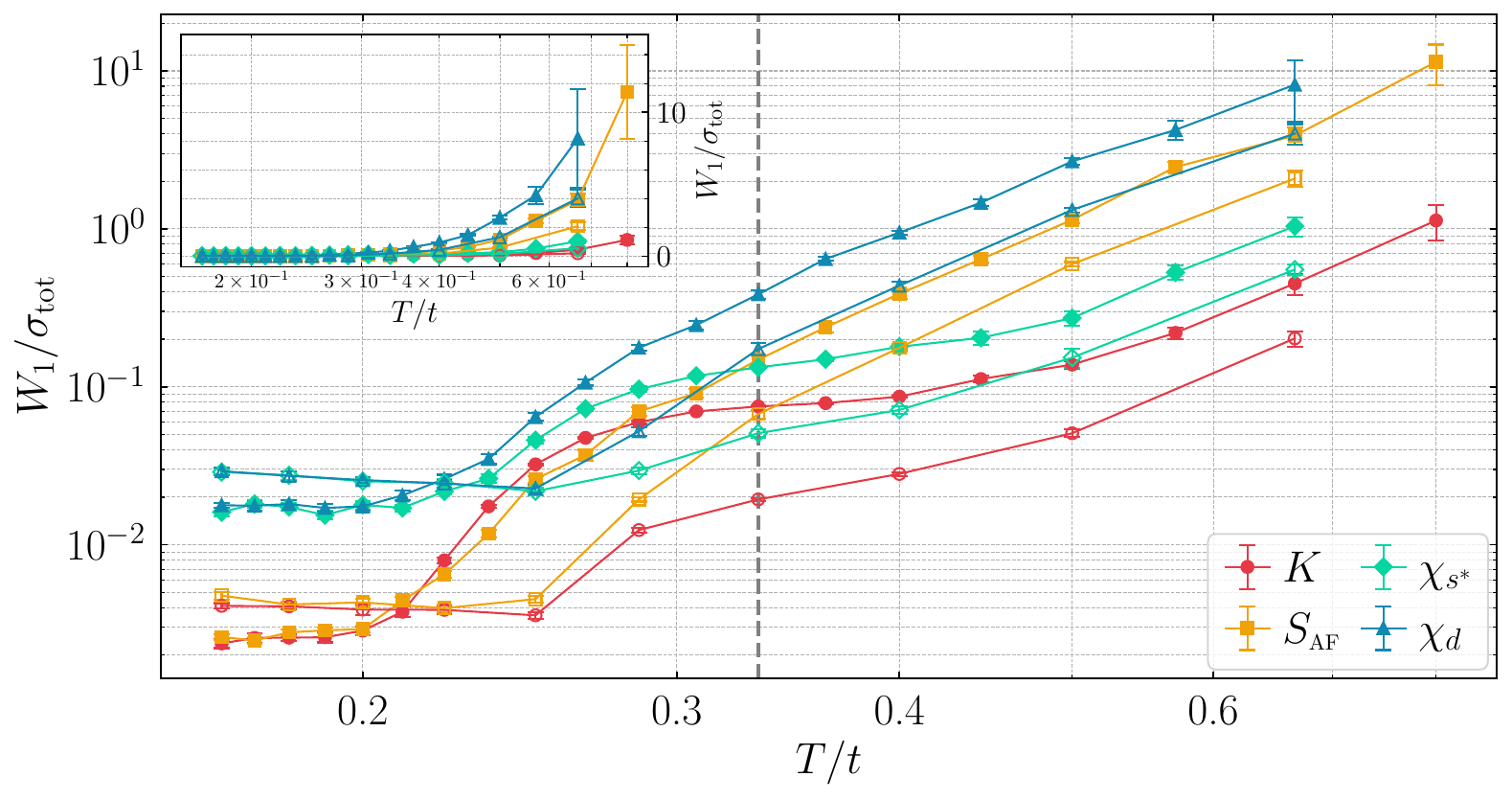}
\caption{The Wasserstein distance $W_1$ between the positive and negative sign histograms at $U/t=6$, normalized by a measure of the standard deviation of the combined distribution of both signs $\sigma_{\rm tot}$; results are averaged over 24 independent Markov chains, and the error bars are the standard error of the means. The normalization leads to a dimensionless quantity that facilitates the comparison of different physical quantities, specifically the ones shown in Figs.~\ref{fig:fig_1_mod} and \ref{fig:fig_2_mod}. The temperature selected in Fig.~\ref{fig:fig_1_mod} is marked as a vertical dashed line. System sizes are $L=8$ (solid markers) and $L=16$ (empty markers) for $\mu = -1.4t$; the inset shows the same data on a linear vertical scale. 
}
\label{fig:fig_3}
\end{figure}

Further quantification of the dissimilarity of the distributions is reported in Fig.~\ref{fig:fig_3}, showing the temperature dependence of the Wasserstein distance (WD) \cite{kantorovich1960mathematical,vaserstein1969markov} between the ${\cal S}=\pm 1$ histograms, a metric which has the useful feature that it can be defined in a way which works directly with the list of measured values, independent of binning --- further details in the SM \cite{SM}, where we also explore a second method of comparing histograms due to Bhattacharyya, with qualitatively similar results. Here, the WD, normalized by the standard deviation of the mixture of distributions of both signs, ${\cal W}_{1}/\sigma_{\rm tot}$, for the four observables in Fig.~\ref{fig:fig_1_mod}, shows that the largest values of the WD between the histograms occur at high-$T$.

This is initially counterintuitive, as the SP is not an issue at this regime, but it reemphasizes that the two distributions effectively sample different parts of phase space. While an explanation in DQMC is elusive, it can be easily understood within the framework of world-line QMC~\cite{hirsch82}. There, the position of the particles at $\tau=\beta$ must be identical to the positions at $\tau=0$, since the partition function ${\cal Z}$ is a trace. The simplest way to obey this requirement is if all the world lines propagate essentially `straight up', i.e., each particle returning to its own initial position. Such a configuration would have ${\cal S}=+1$. An ${\cal S}=-1$ configuration arises when the world lines satisfy the trace requirement by exchanging during the imaginary time propagation. For this to happen at high $T$ (small $\beta$), particles must have a high `velocity'--- changing their positions a significant amount in a short (imaginary) time.  Although the work reported here employs DQMC rather than world-line QMC, this picture suggests that a high temperature ${\cal S}=-1$ configuration would tend to have an atypically large kinetic energy, and consequently, the ${\cal S}=-1$ histogram will be differently shaped from the ${\cal S}=+1$ histogram providing a qualitative explanation for Fig.~\ref{fig:fig_3}. 

\prlsection{Means and the measurement bias} While the difference between reweighted and unreweighted observables such as $\langle \chi_{d}\rangle_{\cal W}-\langle \chi_{d}\rangle_{|{\cal W}|}$ is known to increase as $T$ decreases~\cite{white89a}, the sign-resolved histograms $P_\pm(\chi_d)$ appear to become increasingly similar in shape --- see inset in Fig.~\ref{fig:fig_3} where the WD between sign-resolved histograms is significantly suppressed. Indeed, Fig.~\ref{fig:fig_3} main panel shows that the normalized Wasserstein distance $W_1/\sigma_{\rm tot}$ between the ${\cal S}=\pm1$ histograms decreases as $T$ is lowered but then saturates to a small but finite plateau for $T/t \lesssim 0.2$. This plateau simply indicates that further cooling no longer significantly changes the shape difference between $P_+$ and $P_-$; it does not mean that the two distributions have become identical, nor that the bias must vanish. 

Crucially, the measurement bias is not controlled by $W_1$ directly, but by the mean difference $\Delta\mu = \langle{\cal O}\rangle_+ - \langle{\cal O}\rangle_-$ and the average sign $\langle{\cal S}\rangle$. This can be seen via the expectation values, $\langle {\cal O} \rangle_{\cal W} = \frac{\langle {\cal O} \rangle_+ -\langle {\cal O} \rangle_-}{\langle {\cal S} \rangle}$ [see Eq.~\ref{eq:expect3}], which, for remaining finite, must have $\langle {\cal O}\rangle_+ - {\langle {\cal O}\rangle_-}\to 0$ since $\langle {\cal S}\rangle\to0$ as $T\to0$. This vanishing of the first moments does not contradict the finite $W_1$: higher moments of the distributions can remain distinct, and $W_1$ is sensitive to these global differences. A more explicit connection to the bias is derived in the SM~\cite{SM}, yielding
\begin{align}
\langle {\cal O}\rangle_{\cal W} - \langle{\cal O}\rangle_{|{\cal W}|} 
= \frac{\Delta\mu\,(1-\langle {\cal S}\rangle ^2)}{2\langle {\cal S}\rangle } \, .
\label{eq:sign_no_sign_relation}
\end{align}
This relation resolves the apparent tension between increasingly similar histograms and an increasing discrepancy between reweighted and unreweighted averages: even though $\Delta \mu$ decreases as $T$ is lowered, the amplification factor $(1-\langle{\cal S}\rangle^2)/(2\langle{\cal S}\rangle)$ grows much faster as $\langle{\cal S}\rangle\to 0$, so the overall bias can (and does) grow at low temperatures.

\begin{figure}[t]
\includegraphics[width=1\columnwidth]{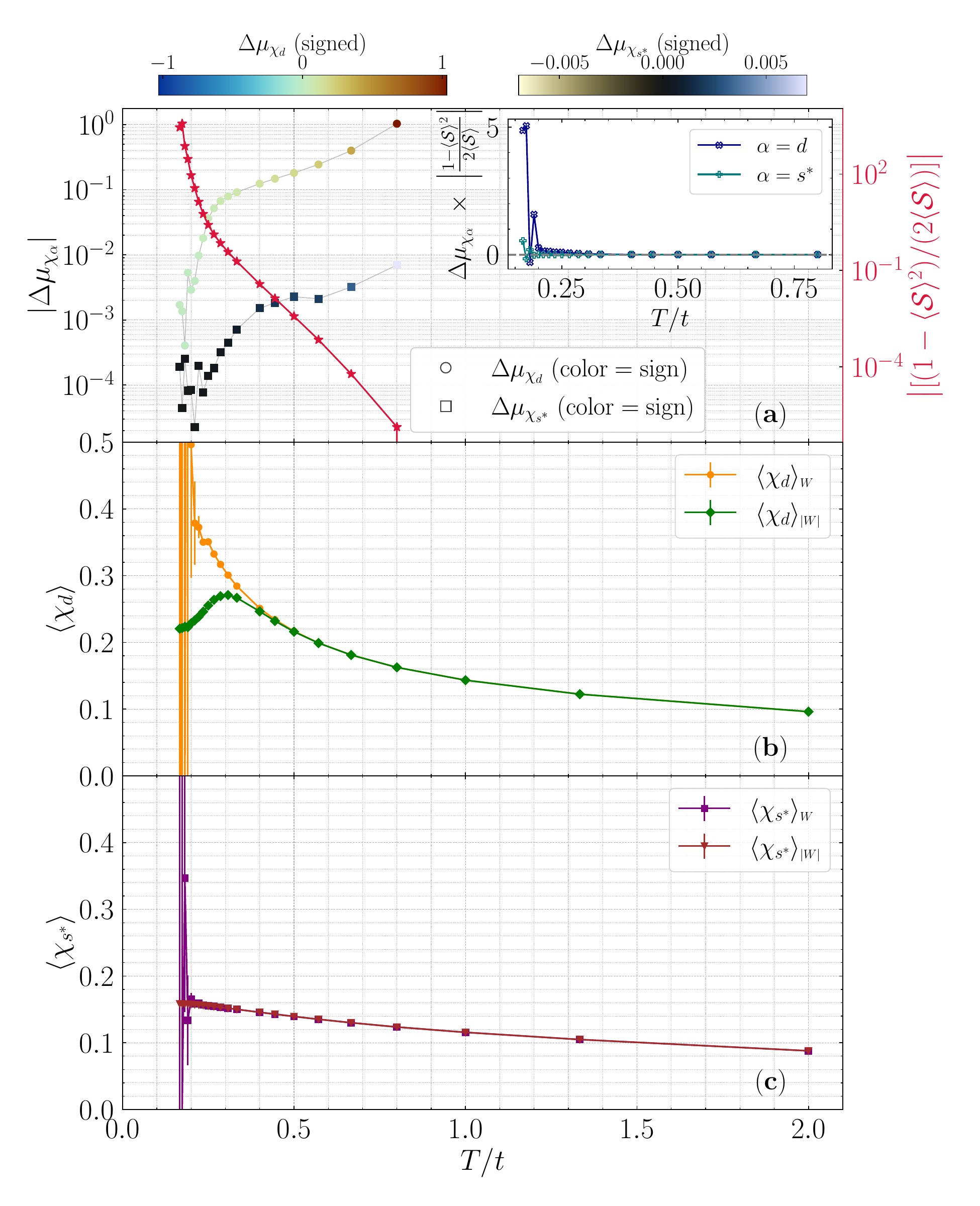}
\caption{(a) The difference between the means $\Delta \mu_{\chi_{\alpha}}$ of the $\alpha = d, s^*$-wave susceptibilities $\chi_d$ and $\chi_{s^*}$ for configurations that have an associated positive and negative weights as a function of temperature; the right axis shows the fraction that enters Eq.~\eqref{eq:sign_no_sign_relation}. (b) [(c)] The average $\chi_d$ [$\chi_{s^*}$] when computing via considering the sign of the weights (i.e., the reweighted average) and ignoring the sign; error bars stem from a jackknife analysis. The inset in (a) shows the right-hand side of Eq.~\eqref{eq:sign_no_sign_relation} for these two quantities, which is proportional to the difference of the curves in (b) and (c). Unlike previous cases, we tune the chemical potential $\mu$ here to achieve a total density $\rho \simeq 0.875$; other parameters are similar to previous figures.}
\label{fig:fig_4}
\end{figure}

Figure~\ref{fig:fig_4}(a) illustrates this compensation explicitly: The decline of $\Delta\mu_{\cal O}$ with decreasing $T$ is accompanied by a rapid growth of the amplification factor, shown on the right axis. Because this factor is independent of the observable, the resulting bias scales directly with the magnitude of $\Delta\mu_{\cal O}$. Since $\Delta\mu_{\chi_d}/\Delta\mu_{\chi_{s^*}}\simeq 10$ at $T\to0$, hence, for the $s^*$-wave susceptibility, the bias remains small, whereas for the $d$-wave susceptibility, it becomes dominant at low temperatures. Figures~\ref{fig:fig_4}(b) and \ref{fig:fig_4}(c) confirm this contrast, showing that neglecting the sign yields a qualitatively incorrect temperature dependence for $\chi_d$, while $\chi_{s^*}$ is only weakly affected.

\prlsection{Summary}
A deeper understanding of the fermion sign problem is crucial for extending the reach of Quantum Monte Carlo simulations of correlated electron systems. In this work, we characterized how the distributions of key observables in the Hubbard model depend on the configuration sign in DQMC, and we established an exact relation linking the bias from neglecting the sign to the difference between sign-resolved means and the average sign $\langle{\cal S}\rangle$. This framework provides a quantitative and intuitive diagnostic of how and when the sign problem alters measured observables.

A striking outcome is that the sign-resolved histograms become nearly identical at low temperature, even as $\langle{\cal S}\rangle \to 0$ and the inclusion of the sign becomes ever more essential. Equation~\eqref{eq:sign_no_sign_relation} explains this paradox: the diminishing difference of means $\Delta\mu$ is outweighed by the diverging amplification factor $[(1-\langle{\cal S}\rangle^2)/(2\langle{\cal S}\rangle)]$, which magnifies the bias of unreweighted observables. This analysis clarifies why certain observables, such as the $d$-wave pairing susceptibility, are quantitatively misrepresented when the sign is ignored, since they exhibit comparatively larger $\Delta\mu$.

Finally, our results pave the way for quantifying the severity of the sign problem through measurable histogram statistics, rather than solely through exponential signal-to-noise decay. An important open question is how the amplification factor and the sign-resolved structure evolve with lattice geometry, the model under investigation, and the type of QMC method used, since the `amplification factor' via $\langle {\cal S} \rangle$ depends on all of them.

\vskip0.10in
\begin{acknowledgments}
R.L.~and R.T.S.~acknowledge support from the U.S.~Department of Energy, Office of Science, Office of Basic Energy Sciences, under Award Number DE-SC0014671. R.M.~acknowledges support from the T$_{\rm c}$SUH Welch Professorship Award. Numerical simulations were performed with resources provided by the Research Computing Data Core at the University of Houston. This work also used TAMU ACES at Texas A\&M HPRC through allocation PHY240046 from the Advanced Cyberinfrastructure Coordination Ecosystem: Services \& Support (ACCESS) program, which is supported by U.S. National Science Foundation grants 2138259, 2138286, 2138307, 2137603, and 2138296. The data that support the findings of this article are openly available at \cite{zenodo}.
\end{acknowledgments}

\bibliography{signresolvedrefs}

\begin{thebibliography}{44}%
\makeatletter
\providecommand \@ifxundefined [1]{%
 \@ifx{#1\undefined}
}%
\providecommand \@ifnum [1]{%
 \ifnum #1\expandafter \@firstoftwo
 \else \expandafter \@secondoftwo
 \fi
}%
\providecommand \@ifx [1]{%
 \ifx #1\expandafter \@firstoftwo
 \else \expandafter \@secondoftwo
 \fi
}%
\providecommand \natexlab [1]{#1}%
\providecommand \enquote  [1]{``#1''}%
\providecommand \bibnamefont  [1]{#1}%
\providecommand \bibfnamefont [1]{#1}%
\providecommand \citenamefont [1]{#1}%
\providecommand \href@noop [0]{\@secondoftwo}%
\providecommand \href [0]{\begingroup \@sanitize@url \@href}%
\providecommand \@href[1]{\@@startlink{#1}\@@href}%
\providecommand \@@href[1]{\endgroup#1\@@endlink}%
\providecommand \@sanitize@url [0]{\catcode `\\12\catcode `\$12\catcode
  `\&12\catcode `\#12\catcode `\^12\catcode `\_12\catcode `\%12\relax}%
\providecommand \@@startlink[1]{}%
\providecommand \@@endlink[0]{}%
\providecommand \url  [0]{\begingroup\@sanitize@url \@url }%
\providecommand \@url [1]{\endgroup\@href {#1}{\urlprefix }}%
\providecommand \urlprefix  [0]{URL }%
\providecommand \Eprint [0]{\href }%
\providecommand \doibase [0]{https://doi.org/}%
\providecommand \selectlanguage [0]{\@gobble}%
\providecommand \bibinfo  [0]{\@secondoftwo}%
\providecommand \bibfield  [0]{\@secondoftwo}%
\providecommand \translation [1]{[#1]}%
\providecommand \BibitemOpen [0]{}%
\providecommand \bibitemStop [0]{}%
\providecommand \bibitemNoStop [0]{.\EOS\space}%
\providecommand \EOS [0]{\spacefactor3000\relax}%
\providecommand \BibitemShut  [1]{\csname bibitem#1\endcsname}%
\let\auto@bib@innerbib\@empty
%</preamble>
\bibitem [{\citenamefont {Hammond}\ \emph {et~al.}(1994)\citenamefont
  {Hammond}, \citenamefont {Lester},\ and\ \citenamefont
  {Reynolds}}]{hammond94}%
  \BibitemOpen
  \bibfield  {author} {\bibinfo {author} {\bibfnamefont {B.}~\bibnamefont
  {Hammond}}, \bibinfo {author} {\bibfnamefont {W.}~\bibnamefont {Lester}},\
  and\ \bibinfo {author} {\bibfnamefont {P.}~\bibnamefont {Reynolds}},\
  }\href@noop {} {\emph {\bibinfo {title} {{Monte Carlo Methods in Ab Initio
  Quantum Chemistry}}}},\ {Lecture and Course Notes in Chemistry: Volume 1}\
  (\bibinfo  {publisher} {World Scientific},\ \bibinfo {year}
  {1994})\BibitemShut {NoStop}%
\bibitem [{\citenamefont {Ceperley}(1995)}]{ceperley95}%
  \BibitemOpen
  \bibfield  {author} {\bibinfo {author} {\bibfnamefont {D.}~\bibnamefont
  {Ceperley}},\ }\bibfield  {title} {\bibinfo {title} {Path integrals in the
  theory of condensed {H}elium},\ }\href
  {https://doi.org/10.1103/RevModPhys.67.279} {\bibfield  {journal} {\bibinfo
  {journal} {Rev. Mod. Phys.}\ }\textbf {\bibinfo {volume} {67}},\ \bibinfo
  {pages} {279} (\bibinfo {year} {1995})}\BibitemShut {NoStop}%
\bibitem [{\citenamefont {Foulkes}\ \emph {et~al.}(2001)\citenamefont
  {Foulkes}, \citenamefont {Mitas}, \citenamefont {Needs},\ and\ \citenamefont
  {Rajagopal}}]{foulkes01}%
  \BibitemOpen
  \bibfield  {author} {\bibinfo {author} {\bibfnamefont {W.}~\bibnamefont
  {Foulkes}}, \bibinfo {author} {\bibfnamefont {L.}~\bibnamefont {Mitas}},
  \bibinfo {author} {\bibfnamefont {R.}~\bibnamefont {Needs}},\ and\ \bibinfo
  {author} {\bibfnamefont {G.}~\bibnamefont {Rajagopal}},\ }\bibfield  {title}
  {\bibinfo {title} {{Quantum Monte Carlo simulations of solids}},\ }\href
  {https://doi.org/10.1103/RevModPhys.73.33} {\bibfield  {journal} {\bibinfo
  {journal} {Rev. Mod. Phys.}\ }\textbf {\bibinfo {volume} {73}},\ \bibinfo
  {pages} {33} (\bibinfo {year} {2001})}\BibitemShut {NoStop}%
\bibitem [{\citenamefont {Degrand}\ and\ \citenamefont
  {DeTar}(2006)}]{degrand06}%
  \BibitemOpen
  \bibfield  {author} {\bibinfo {author} {\bibfnamefont {T.}~\bibnamefont
  {Degrand}}\ and\ \bibinfo {author} {\bibfnamefont {C.}~\bibnamefont
  {DeTar}},\ }\href {https://doi.org/https://doi.org/10.1142/6065} {\emph
  {\bibinfo {title} {Lattice Methods for Quantum Chromodynamics}}}\ (\bibinfo
  {publisher} {World Scientific},\ \bibinfo {year} {2006})\BibitemShut
  {NoStop}%
\bibitem [{\citenamefont {Carlson}\ \emph {et~al.}(2015)\citenamefont
  {Carlson}, \citenamefont {Gandolfi}, \citenamefont {Pederiva}, \citenamefont
  {Pieper}, \citenamefont {Schiavilla}, \citenamefont {Schmidt},\ and\
  \citenamefont {Wiringa}}]{carlson15}%
  \BibitemOpen
  \bibfield  {author} {\bibinfo {author} {\bibfnamefont {J.}~\bibnamefont
  {Carlson}}, \bibinfo {author} {\bibfnamefont {S.}~\bibnamefont {Gandolfi}},
  \bibinfo {author} {\bibfnamefont {F.}~\bibnamefont {Pederiva}}, \bibinfo
  {author} {\bibfnamefont {S.~C.}\ \bibnamefont {Pieper}}, \bibinfo {author}
  {\bibfnamefont {R.}~\bibnamefont {Schiavilla}}, \bibinfo {author}
  {\bibfnamefont {K.}~\bibnamefont {Schmidt}},\ and\ \bibinfo {author}
  {\bibfnamefont {R.}~\bibnamefont {Wiringa}},\ }\bibfield  {title} {\bibinfo
  {title} {{Quantum Monte Carlo methods for nuclear physics}},\ }\href
  {https://doi.org/10.1103/RevModPhys.87.1067} {\bibfield  {journal} {\bibinfo
  {journal} {Rev. Mod. Phys.}\ }\textbf {\bibinfo {volume} {87}},\ \bibinfo
  {pages} {1067} (\bibinfo {year} {2015})}\BibitemShut {NoStop}%
\bibitem [{\citenamefont {Needs}\ \emph {et~al.}(2020)\citenamefont {Needs},
  \citenamefont {Towler}, \citenamefont {Drummond}, \citenamefont
  {L\'opez~R\'ios},\ and\ \citenamefont {Trail}}]{needs20}%
  \BibitemOpen
  \bibfield  {author} {\bibinfo {author} {\bibfnamefont {R.}~\bibnamefont
  {Needs}}, \bibinfo {author} {\bibfnamefont {M.}~\bibnamefont {Towler}},
  \bibinfo {author} {\bibfnamefont {N.}~\bibnamefont {Drummond}}, \bibinfo
  {author} {\bibfnamefont {P.}~\bibnamefont {L\'opez~R\'ios}},\ and\ \bibinfo
  {author} {\bibfnamefont {J.}~\bibnamefont {Trail}},\ }\bibfield  {title}
  {\bibinfo {title} {Variational and diffusion quantum {M}onte {C}arlo
  calculations with the {CASINO} code},\ }\href
  {https://doi.org/10.1063/1.5144288} {\bibfield  {journal} {\bibinfo
  {journal} {J. Chem. Phys.}\ }\textbf {\bibinfo {volume} {152}},\ \bibinfo
  {pages} {154106} (\bibinfo {year} {2020})}\BibitemShut {NoStop}%
\bibitem [{\citenamefont {Troyer}\ and\ \citenamefont
  {Wiese}(2005)}]{troyer05}%
  \BibitemOpen
  \bibfield  {author} {\bibinfo {author} {\bibfnamefont {M.}~\bibnamefont
  {Troyer}}\ and\ \bibinfo {author} {\bibfnamefont {U.-J.}\ \bibnamefont
  {Wiese}},\ }\bibfield  {title} {\bibinfo {title} {Computational complexity
  and fundamental limitations to fermionic quantum {M}onte {C}arlo
  simulations},\ }\href {https://doi.org/10.1103/PhysRevLett.94.170201}
  {\bibfield  {journal} {\bibinfo  {journal} {Phys. Rev. Lett.}\ }\textbf
  {\bibinfo {volume} {94}},\ \bibinfo {pages} {170201} (\bibinfo {year}
  {2005})}\BibitemShut {NoStop}%
\bibitem [{\citenamefont {Loh}\ \emph {et~al.}(1990)\citenamefont {Loh},
  \citenamefont {Gubernatis}, \citenamefont {Scalettar}, \citenamefont {White},
  \citenamefont {Scalapino},\ and\ \citenamefont {Sugar}}]{loh90}%
  \BibitemOpen
  \bibfield  {author} {\bibinfo {author} {\bibfnamefont {E.}~\bibnamefont
  {Loh}}, \bibinfo {author} {\bibfnamefont {J.}~\bibnamefont {Gubernatis}},
  \bibinfo {author} {\bibfnamefont {R.}~\bibnamefont {Scalettar}}, \bibinfo
  {author} {\bibfnamefont {S.}~\bibnamefont {White}}, \bibinfo {author}
  {\bibfnamefont {D.}~\bibnamefont {Scalapino}},\ and\ \bibinfo {author}
  {\bibfnamefont {R.}~\bibnamefont {Sugar}},\ }\bibfield  {title} {\bibinfo
  {title} {Sign problem in the numerical simulation of many-electron systems},\
  }\href {https://doi.org/10.1103/PhysRevB.41.9301} {\bibfield  {journal}
  {\bibinfo  {journal} {Phys. Rev. B}\ }\textbf {\bibinfo {volume} {41}},\
  \bibinfo {pages} {9301} (\bibinfo {year} {1990})}\BibitemShut {NoStop}%
\bibitem [{\citenamefont {Iglovikov}\ \emph {et~al.}(2015)\citenamefont
  {Iglovikov}, \citenamefont {Khatami},\ and\ \citenamefont
  {Scalettar}}]{iglovikov15}%
  \BibitemOpen
  \bibfield  {author} {\bibinfo {author} {\bibfnamefont {V.}~\bibnamefont
  {Iglovikov}}, \bibinfo {author} {\bibfnamefont {E.}~\bibnamefont {Khatami}},\
  and\ \bibinfo {author} {\bibfnamefont {R.}~\bibnamefont {Scalettar}},\
  }\bibfield  {title} {\bibinfo {title} {{Geometry dependence of the sign
  problem in quantum Monte Carlo simulations}},\ }\href
  {https://doi.org/10.1103/PhysRevB.92.045110} {\bibfield  {journal} {\bibinfo
  {journal} {Phys. Rev. B}\ }\textbf {\bibinfo {volume} {92}},\ \bibinfo
  {pages} {045110} (\bibinfo {year} {2015})}\BibitemShut {NoStop}%
\bibitem [{\citenamefont {Wu}\ and\ \citenamefont {Zhang}(2005)}]{wu05}%
  \BibitemOpen
  \bibfield  {author} {\bibinfo {author} {\bibfnamefont {C.}~\bibnamefont
  {Wu}}\ and\ \bibinfo {author} {\bibfnamefont {S.-C.}\ \bibnamefont {Zhang}},\
  }\bibfield  {title} {\bibinfo {title} {{Sufficient condition for absence of
  the sign problem in the fermionic quantum Monte Carlo algorithm}},\ }\href
  {https://doi.org/10.1103/PhysRevB.71.155115} {\bibfield  {journal} {\bibinfo
  {journal} {Phys. Rev. B}\ }\textbf {\bibinfo {volume} {71}},\ \bibinfo
  {pages} {155115} (\bibinfo {year} {2005})}\BibitemShut {NoStop}%
\bibitem [{\citenamefont {Li}\ \emph {et~al.}(2016)\citenamefont {Li},
  \citenamefont {Jiang},\ and\ \citenamefont {Yao}}]{li16}%
  \BibitemOpen
  \bibfield  {author} {\bibinfo {author} {\bibfnamefont {Z.-X.}\ \bibnamefont
  {Li}}, \bibinfo {author} {\bibfnamefont {Y.-F.}\ \bibnamefont {Jiang}},\ and\
  \bibinfo {author} {\bibfnamefont {H.}~\bibnamefont {Yao}},\ }\bibfield
  {title} {\bibinfo {title} {Majorana-time-reversal symmetries: A fundamental
  principle for sign-problem-free quantum {M}onte {C}arlo simulations},\ }\href
  {https://doi.org/10.1103/PhysRevLett.117.267002} {\bibfield  {journal}
  {\bibinfo  {journal} {Phys. Rev. Lett.}\ }\textbf {\bibinfo {volume} {117}},\
  \bibinfo {pages} {267002} (\bibinfo {year} {2016})}\BibitemShut {NoStop}%
\bibitem [{\citenamefont {Chandrasekharan}(2010)}]{chandrasekharan10}%
  \BibitemOpen
  \bibfield  {author} {\bibinfo {author} {\bibfnamefont {S.}~\bibnamefont
  {Chandrasekharan}},\ }\bibfield  {title} {\bibinfo {title} {Fermion bag
  approach to lattice field theories},\ }\href
  {https://doi.org/10.1103/PhysRevD.82.025007} {\bibfield  {journal} {\bibinfo
  {journal} {Phys. Rev. D}\ }\textbf {\bibinfo {volume} {82}},\ \bibinfo
  {pages} {025007} (\bibinfo {year} {2010})}\BibitemShut {NoStop}%
\bibitem [{\citenamefont {Berg}\ \emph {et~al.}(2012)\citenamefont {Berg},
  \citenamefont {Metlitski},\ and\ \citenamefont {Sachdev}}]{berg12}%
  \BibitemOpen
  \bibfield  {author} {\bibinfo {author} {\bibfnamefont {E.}~\bibnamefont
  {Berg}}, \bibinfo {author} {\bibfnamefont {M.~A.}\ \bibnamefont
  {Metlitski}},\ and\ \bibinfo {author} {\bibfnamefont {S.}~\bibnamefont
  {Sachdev}},\ }\bibfield  {title} {\bibinfo {title}
  {Sign-problem{\textendash}free quantum {M}onte {C}arlo of the onset of
  antiferromagnetism in metals},\ }\href
  {https://doi.org/10.1126/science.1227769} {\bibfield  {journal} {\bibinfo
  {journal} {Science}\ }\textbf {\bibinfo {volume} {338}},\ \bibinfo {pages}
  {1606} (\bibinfo {year} {2012})}\BibitemShut {NoStop}%
\bibitem [{\citenamefont {Wang}\ \emph {et~al.}(2015)\citenamefont {Wang},
  \citenamefont {Liu}, \citenamefont {Iazzi}, \citenamefont {Troyer},\ and\
  \citenamefont {Harcos}}]{wang15}%
  \BibitemOpen
  \bibfield  {author} {\bibinfo {author} {\bibfnamefont {L.}~\bibnamefont
  {Wang}}, \bibinfo {author} {\bibfnamefont {Y.-H.}\ \bibnamefont {Liu}},
  \bibinfo {author} {\bibfnamefont {M.}~\bibnamefont {Iazzi}}, \bibinfo
  {author} {\bibfnamefont {M.}~\bibnamefont {Troyer}},\ and\ \bibinfo {author}
  {\bibfnamefont {G.}~\bibnamefont {Harcos}},\ }\bibfield  {title} {\bibinfo
  {title} {Split orthogonal group: A guiding principle for sign-problem-free
  fermionic simulations},\ }\href
  {https://doi.org/10.1103/PhysRevLett.115.250601} {\bibfield  {journal}
  {\bibinfo  {journal} {Phys. Rev. Lett.}\ }\textbf {\bibinfo {volume} {115}},\
  \bibinfo {pages} {250601} (\bibinfo {year} {2015})}\BibitemShut {NoStop}%
\bibitem [{\citenamefont {Li}\ and\ \citenamefont {Yao}(2019)}]{li19}%
  \BibitemOpen
  \bibfield  {author} {\bibinfo {author} {\bibfnamefont {Z.-X.}\ \bibnamefont
  {Li}}\ and\ \bibinfo {author} {\bibfnamefont {H.}~\bibnamefont {Yao}},\
  }\bibfield  {title} {\bibinfo {title} {Sign-problem-free fermionic quantum
  {M}onte {C}arlo: Developments and applications},\ }\href
  {https://doi.org/10.1146/annurev-conmatphys-033117-054307} {\bibfield
  {journal} {\bibinfo  {journal} {Annu. Rev. Condens. Matter Phys.}\ }\textbf
  {\bibinfo {volume} {10}},\ \bibinfo {pages} {337} (\bibinfo {year}
  {2019})}\BibitemShut {NoStop}%
\bibitem [{\citenamefont {Wang}\ \emph {et~al.}(2014)\citenamefont {Wang},
  \citenamefont {Corboz},\ and\ \citenamefont {Troyer}}]{Wang2014}%
  \BibitemOpen
  \bibfield  {author} {\bibinfo {author} {\bibfnamefont {L.}~\bibnamefont
  {Wang}}, \bibinfo {author} {\bibfnamefont {P.}~\bibnamefont {Corboz}},\ and\
  \bibinfo {author} {\bibfnamefont {M.}~\bibnamefont {Troyer}},\ }\bibfield
  {title} {\bibinfo {title} {Fermionic quantum critical point of spinless
  fermions on a honeycomb lattice},\ }\href
  {https://doi.org/10.1088/1367-2630/16/10/103008} {\bibfield  {journal}
  {\bibinfo  {journal} {New J. Phys.}\ }\textbf {\bibinfo {volume} {16}},\
  \bibinfo {pages} {103008} (\bibinfo {year} {2014})}\BibitemShut {NoStop}%
\bibitem [{\citenamefont {Li}\ \emph {et~al.}(2015)\citenamefont {Li},
  \citenamefont {Jiang},\ and\ \citenamefont {Yao}}]{ZiXiang2015}%
  \BibitemOpen
  \bibfield  {author} {\bibinfo {author} {\bibfnamefont {Z.-X.}\ \bibnamefont
  {Li}}, \bibinfo {author} {\bibfnamefont {Y.-F.}\ \bibnamefont {Jiang}},\ and\
  \bibinfo {author} {\bibfnamefont {H.}~\bibnamefont {Yao}},\ }\bibfield
  {title} {\bibinfo {title} {{Solving the fermion sign problem in quantum Monte
  Carlo simulations by Majorana representation}},\ }\href
  {https://doi.org/10.1103/PhysRevB.91.241117} {\bibfield  {journal} {\bibinfo
  {journal} {Phys. Rev. B}\ }\textbf {\bibinfo {volume} {91}},\ \bibinfo
  {pages} {241117} (\bibinfo {year} {2015})}\BibitemShut {NoStop}%
\bibitem [{\citenamefont {Hangleiter}\ \emph {et~al.}(2020)\citenamefont
  {Hangleiter}, \citenamefont {Roth}, \citenamefont {Nagaj},\ and\
  \citenamefont {Eisert}}]{Hangleiter2020}%
  \BibitemOpen
  \bibfield  {author} {\bibinfo {author} {\bibfnamefont {D.}~\bibnamefont
  {Hangleiter}}, \bibinfo {author} {\bibfnamefont {I.}~\bibnamefont {Roth}},
  \bibinfo {author} {\bibfnamefont {D.}~\bibnamefont {Nagaj}},\ and\ \bibinfo
  {author} {\bibfnamefont {J.}~\bibnamefont {Eisert}},\ }\bibfield  {title}
  {\bibinfo {title} {Easing the {M}onte {C}arlo sign problem},\ }\bibfield
  {journal} {\bibinfo  {journal} {Science Advances}\ }\textbf {\bibinfo
  {volume} {6}},\ \href {https://doi.org/10.1126/sciadv.abb8341}
  {10.1126/sciadv.abb8341} (\bibinfo {year} {2020})\BibitemShut {NoStop}%
\bibitem [{\citenamefont {Wan}\ \emph {et~al.}(2020)\citenamefont {Wan},
  \citenamefont {Zhang},\ and\ \citenamefont {Yao}}]{Wan2020}%
  \BibitemOpen
  \bibfield  {author} {\bibinfo {author} {\bibfnamefont {Z.-Q.}\ \bibnamefont
  {Wan}}, \bibinfo {author} {\bibfnamefont {S.-X.}\ \bibnamefont {Zhang}},\
  and\ \bibinfo {author} {\bibfnamefont {H.}~\bibnamefont {Yao}},\ }\href@noop
  {} {\bibinfo {title} {Mitigating sign problem by automatic differentiation}}
  (\bibinfo {year} {2020}),\ \Eprint {https://arxiv.org/abs/2010.01141}
  {arXiv:2010.01141 [cond-mat.str-el]} \BibitemShut {NoStop}%
\bibitem [{\citenamefont {Wessel}\ \emph {et~al.}(7 18)\citenamefont {Wessel},
  \citenamefont {Normand}, \citenamefont {Mila},\ and\ \citenamefont
  {Honecker}}]{wessel2017efficient}%
  \BibitemOpen
  \bibfield  {author} {\bibinfo {author} {\bibfnamefont {S.}~\bibnamefont
  {Wessel}}, \bibinfo {author} {\bibfnamefont {B.}~\bibnamefont {Normand}},
  \bibinfo {author} {\bibfnamefont {F.}~\bibnamefont {Mila}},\ and\ \bibinfo
  {author} {\bibfnamefont {A.}~\bibnamefont {Honecker}},\ }\bibfield  {title}
  {\bibinfo {title} {Efficient quantum {M}onte {C}arlo simulations of highly
  frustrated magnets: the frustrated spin-1/2 ladder},\ }\href
  {https://scipost.org/10.21468/SciPostPhys.3.1.005/pdf} {\bibfield  {journal}
  {\bibinfo  {journal} {SciPost Physics.}\ }\textbf {\bibinfo {volume} {3}}
  (\bibinfo {year} {2017-07-18})}\BibitemShut {NoStop}%
\bibitem [{\citenamefont {Mondaini}\ \emph {et~al.}(2022)\citenamefont
  {Mondaini}, \citenamefont {Tarat},\ and\ \citenamefont
  {Scalettar}}]{mondaini22}%
  \BibitemOpen
  \bibfield  {author} {\bibinfo {author} {\bibfnamefont {R.}~\bibnamefont
  {Mondaini}}, \bibinfo {author} {\bibfnamefont {S.}~\bibnamefont {Tarat}},\
  and\ \bibinfo {author} {\bibfnamefont {R.~T.}\ \bibnamefont {Scalettar}},\
  }\bibfield  {title} {\bibinfo {title} {Quantum critical points and the sign
  problem},\ }\href {https://doi.org/10.1126/science.abg9299} {\bibfield
  {journal} {\bibinfo  {journal} {Science}\ }\textbf {\bibinfo {volume}
  {375}},\ \bibinfo {pages} {418} (\bibinfo {year} {2022})}\BibitemShut
  {NoStop}%
\bibitem [{\citenamefont {Mou}\ \emph {et~al.}(2022)\citenamefont {Mou},
  \citenamefont {Mondaini},\ and\ \citenamefont {Scalettar}}]{mou22}%
  \BibitemOpen
  \bibfield  {author} {\bibinfo {author} {\bibfnamefont {Y.}~\bibnamefont
  {Mou}}, \bibinfo {author} {\bibfnamefont {R.}~\bibnamefont {Mondaini}},\ and\
  \bibinfo {author} {\bibfnamefont {R.~T.}\ \bibnamefont {Scalettar}},\
  }\bibfield  {title} {\bibinfo {title} {Bilayer {H}ubbard model: Analysis
  based on the fermionic sign problem},\ }\href
  {https://doi.org/10.1103/PhysRevB.106.125116} {\bibfield  {journal} {\bibinfo
   {journal} {Phys. Rev. B}\ }\textbf {\bibinfo {volume} {106}},\ \bibinfo
  {pages} {125116} (\bibinfo {year} {2022})}\BibitemShut {NoStop}%
\bibitem [{\citenamefont {Mondaini}\ \emph {et~al.}(2023)\citenamefont
  {Mondaini}, \citenamefont {Tarat},\ and\ \citenamefont
  {Scalettar}}]{mondaini23}%
  \BibitemOpen
  \bibfield  {author} {\bibinfo {author} {\bibfnamefont {R.}~\bibnamefont
  {Mondaini}}, \bibinfo {author} {\bibfnamefont {S.}~\bibnamefont {Tarat}},\
  and\ \bibinfo {author} {\bibfnamefont {R.~T.}\ \bibnamefont {Scalettar}},\
  }\bibfield  {title} {\bibinfo {title} {Universality and critical exponents of
  the fermion sign problem},\ }\href
  {https://doi.org/10.1103/PhysRevB.107.245144} {\bibfield  {journal} {\bibinfo
   {journal} {Phys. Rev. B}\ }\textbf {\bibinfo {volume} {107}},\ \bibinfo
  {pages} {245144} (\bibinfo {year} {2023})}\BibitemShut {NoStop}%
\bibitem [{\citenamefont {Yi}\ \emph {et~al.}(2024)\citenamefont {Yi},
  \citenamefont {Cheng}, \citenamefont {Pil\'e}, \citenamefont {Burovski},\
  and\ \citenamefont {Mondaini}}]{tiancheng24}%
  \BibitemOpen
  \bibfield  {author} {\bibinfo {author} {\bibfnamefont {T.-C.}\ \bibnamefont
  {Yi}}, \bibinfo {author} {\bibfnamefont {S.}~\bibnamefont {Cheng}}, \bibinfo
  {author} {\bibfnamefont {I.}~\bibnamefont {Pil\'e}}, \bibinfo {author}
  {\bibfnamefont {E.}~\bibnamefont {Burovski}},\ and\ \bibinfo {author}
  {\bibfnamefont {R.}~\bibnamefont {Mondaini}},\ }\bibfield  {title} {\bibinfo
  {title} {Two-dimensional polarized superfluids through the prism of the
  fermion sign problem},\ }\href {https://doi.org/10.1103/PhysRevB.110.085131}
  {\bibfield  {journal} {\bibinfo  {journal} {Phys. Rev. B}\ }\textbf {\bibinfo
  {volume} {110}},\ \bibinfo {pages} {085131} (\bibinfo {year}
  {2024})}\BibitemShut {NoStop}%
\bibitem [{\citenamefont {Hirsch}\ \emph {et~al.}(1982)\citenamefont {Hirsch},
  \citenamefont {Sugar}, \citenamefont {Scalapino},\ and\ \citenamefont
  {Blankenbecler}}]{hirsch82}%
  \BibitemOpen
  \bibfield  {author} {\bibinfo {author} {\bibfnamefont {J.}~\bibnamefont
  {Hirsch}}, \bibinfo {author} {\bibfnamefont {R.}~\bibnamefont {Sugar}},
  \bibinfo {author} {\bibfnamefont {D.}~\bibnamefont {Scalapino}},\ and\
  \bibinfo {author} {\bibfnamefont {R.}~\bibnamefont {Blankenbecler}},\
  }\bibfield  {title} {\bibinfo {title} {Monte {C}arlo simulations of
  one-dimensional fermion systems},\ }\href
  {https://doi.org/10.1103/PhysRevB.26.5033} {\bibfield  {journal} {\bibinfo
  {journal} {Phys. Rev. B}\ }\textbf {\bibinfo {volume} {26}},\ \bibinfo
  {pages} {5033} (\bibinfo {year} {1982})}\BibitemShut {NoStop}%
\bibitem [{\citenamefont {Batrouni}\ and\ \citenamefont
  {de~Forcrand}(1993)}]{batrouni93}%
  \BibitemOpen
  \bibfield  {author} {\bibinfo {author} {\bibfnamefont {G.~G.}\ \bibnamefont
  {Batrouni}}\ and\ \bibinfo {author} {\bibfnamefont {P.}~\bibnamefont
  {de~Forcrand}},\ }\bibfield  {title} {\bibinfo {title} {Fermion sign problem:
  Decoupling transformation and simulation algorithm},\ }\href
  {https://doi.org/10.1103/PhysRevB.48.589} {\bibfield  {journal} {\bibinfo
  {journal} {Phys. Rev. B}\ }\textbf {\bibinfo {volume} {48}},\ \bibinfo
  {pages} {589} (\bibinfo {year} {1993})}\BibitemShut {NoStop}%
\bibitem [{Note1()}]{Note1}%
  \BibitemOpen
  \bibinfo {note} {In models of interacting electrons and phonons, such as the
  Holstein Hamiltonian, the phonon kinetic energy $\protect \hat p_{i}^{2}$
  controls the imaginary time fluctuations of the phonon field, leading to a
  substantial reduction in the presence of configurations for which the fermion
  determinant is negative.}\BibitemShut {Stop}%
\bibitem [{\citenamefont {Blankenbecler}\ \emph {et~al.}(1981)\citenamefont
  {Blankenbecler}, \citenamefont {Scalapino},\ and\ \citenamefont
  {Sugar}}]{blankenbecler81}%
  \BibitemOpen
  \bibfield  {author} {\bibinfo {author} {\bibfnamefont {R.}~\bibnamefont
  {Blankenbecler}}, \bibinfo {author} {\bibfnamefont {D.}~\bibnamefont
  {Scalapino}},\ and\ \bibinfo {author} {\bibfnamefont {R.}~\bibnamefont
  {Sugar}},\ }\bibfield  {title} {\bibinfo {title} {{Monte Carlo calculations
  of coupled boson-fermion systems. I}},\ }\href
  {https://doi.org/10.1103/PhysRevD.24.2278} {\bibfield  {journal} {\bibinfo
  {journal} {Phys. Rev. D}\ }\textbf {\bibinfo {volume} {24}},\ \bibinfo
  {pages} {2278} (\bibinfo {year} {1981})}\BibitemShut {NoStop}%
\bibitem [{\citenamefont {Scalapino}(1994)}]{scalapino94}%
  \BibitemOpen
  \bibfield  {author} {\bibinfo {author} {\bibfnamefont {D.}~\bibnamefont
  {Scalapino}},\ }\bibfield  {title} {\bibinfo {title} {Does the {H}ubbard
  model have the right stuff?},\ }in\ \href@noop {} {\emph {\bibinfo
  {booktitle} {Proceedings of the International School of Physics}}},\ \bibinfo
  {editor} {edited by\ \bibinfo {editor} {\bibfnamefont {R.}~\bibnamefont
  {Broglia}}\ and\ \bibinfo {editor} {\bibfnamefont {J.}~\bibnamefont
  {Schrieffer}}}\ (\bibinfo  {publisher} {North-Holland},\ \bibinfo {year}
  {1994})\BibitemShut {NoStop}%
\bibitem [{\citenamefont {Scalapino}(2012)}]{Scalapino2012}%
  \BibitemOpen
  \bibfield  {author} {\bibinfo {author} {\bibfnamefont {D.~J.}\ \bibnamefont
  {Scalapino}},\ }\bibfield  {title} {\bibinfo {title} {A common thread: {T}he
  pairing interaction for unconventional superconductors},\ }\href
  {https://doi.org/10.1103/RevModPhys.84.1383} {\bibfield  {journal} {\bibinfo
  {journal} {Rev. Mod. Phys.}\ }\textbf {\bibinfo {volume} {84}},\ \bibinfo
  {pages} {1383} (\bibinfo {year} {2012})}\BibitemShut {NoStop}%
\bibitem [{\citenamefont {White}\ \emph {et~al.}(1989)\citenamefont {White},
  \citenamefont {Scalapino}, \citenamefont {Sugar}, \citenamefont {Bickers},\
  and\ \citenamefont {Scalettar}}]{white89a}%
  \BibitemOpen
  \bibfield  {author} {\bibinfo {author} {\bibfnamefont {S.~R.}\ \bibnamefont
  {White}}, \bibinfo {author} {\bibfnamefont {D.~J.}\ \bibnamefont
  {Scalapino}}, \bibinfo {author} {\bibfnamefont {R.~L.}\ \bibnamefont
  {Sugar}}, \bibinfo {author} {\bibfnamefont {N.~E.}\ \bibnamefont {Bickers}},\
  and\ \bibinfo {author} {\bibfnamefont {R.~T.}\ \bibnamefont {Scalettar}},\
  }\bibfield  {title} {\bibinfo {title} {Attractive and repulsive pairing
  interaction vertices for the two-dimensional {H}ubbard model},\ }\href
  {https://doi.org/10.1103/PhysRevB.39.839} {\bibfield  {journal} {\bibinfo
  {journal} {Phys. Rev. B}\ }\textbf {\bibinfo {volume} {39}},\ \bibinfo
  {pages} {839} (\bibinfo {year} {1989})}\BibitemShut {NoStop}%
\bibitem [{\citenamefont {Montorsi}(1992)}]{montorsi1992hubbard}%
  \BibitemOpen
  \bibfield  {author} {\bibinfo {author} {\bibfnamefont {A.}~\bibnamefont
  {Montorsi}},\ }\href
  {https://books.google.com/books?hl=en&lr=&id=QoGl8H7tphYC&oi=fnd&pg=PA1&dq=montorsi+hubbard+model&ots=wAFChwYjCh&sig=WdIHeO0fXy7yfenQ5ZF63Fm2IcM#v=onepage&q=montorsi%20hubbard%20model&f=false}
  {\emph {\bibinfo {title} {The {H}ubbard Model: A Reprint Volume}}}\ (\bibinfo
   {publisher} {World Scientific},\ \bibinfo {year} {1992})\BibitemShut
  {NoStop}%
\bibitem [{\citenamefont {Tasaki}(1998)}]{tasaki1998hubbard}%
  \BibitemOpen
  \bibfield  {author} {\bibinfo {author} {\bibfnamefont {H.}~\bibnamefont
  {Tasaki}},\ }\bibfield  {title} {\bibinfo {title} {The {H}ubbard model-an
  introduction and selected rigorous results},\ }\href
  {https://iopscience.iop.org/article/10.1088/0953-8984/10/20/004/pdf}
  {\bibfield  {journal} {\bibinfo  {journal} {Journal of Physics: Condensed
  Matter}\ }\textbf {\bibinfo {volume} {10}},\ \bibinfo {pages} {4353}
  (\bibinfo {year} {1998})}\BibitemShut {NoStop}%
\bibitem [{\citenamefont {Fazekas}(1999)}]{fazekas1999lecture}%
  \BibitemOpen
  \bibfield  {author} {\bibinfo {author} {\bibfnamefont {P.}~\bibnamefont
  {Fazekas}},\ }\href {https://core.ac.uk/download/pdf/25218749.pdf} {\emph
  {\bibinfo {title} {Lecture notes on electron correlation and magnetism}}},\
  Vol.~\bibinfo {volume} {5}\ (\bibinfo  {publisher} {World scientific},\
  \bibinfo {year} {1999})\BibitemShut {NoStop}%
\bibitem [{\citenamefont {Arovas}\ \emph {et~al.}(2022)\citenamefont {Arovas},
  \citenamefont {Berg}, \citenamefont {Kivelson},\ and\ \citenamefont
  {Raghu}}]{arovas2022hubbard}%
  \BibitemOpen
  \bibfield  {author} {\bibinfo {author} {\bibfnamefont {D.~P.}\ \bibnamefont
  {Arovas}}, \bibinfo {author} {\bibfnamefont {E.}~\bibnamefont {Berg}},
  \bibinfo {author} {\bibfnamefont {S.~A.}\ \bibnamefont {Kivelson}},\ and\
  \bibinfo {author} {\bibfnamefont {S.}~\bibnamefont {Raghu}},\ }\bibfield
  {title} {\bibinfo {title} {The {H}ubbard model},\ }\href
  {https://www.annualreviews.org/content/journals/10.1146/annurev-conmatphys-031620-102024}
  {\bibfield  {journal} {\bibinfo  {journal} {Annual review of condensed matter
  physics}\ }\textbf {\bibinfo {volume} {13}},\ \bibinfo {pages} {239}
  (\bibinfo {year} {2022})}\BibitemShut {NoStop}%
\bibitem [{\citenamefont {Hirsch}(1985)}]{Hirsch1985}%
  \BibitemOpen
  \bibfield  {author} {\bibinfo {author} {\bibfnamefont {J.~E.}\ \bibnamefont
  {Hirsch}},\ }\bibfield  {title} {\bibinfo {title} {Two-dimensional {H}ubbard
  model: Numerical simulation study},\ }\href
  {https://doi.org/10.1103/PhysRevB.31.4403} {\bibfield  {journal} {\bibinfo
  {journal} {Phys. Rev. B}\ }\textbf {\bibinfo {volume} {31}},\ \bibinfo
  {pages} {4403} (\bibinfo {year} {1985})}\BibitemShut {NoStop}%
\bibitem [{Note2()}]{Note2}%
  \BibitemOpen
  \bibinfo {note} {In more refined treatments relevant to the physics of the
  cuprates, a next-near-neighbor hopping $t^{\prime }\simeq -0.2 t$ is often
  included in the hole-doped regime~\cite {huang2018stripe,
  Xu2024}.}\BibitemShut {Stop}%
\bibitem [{SM()}]{SM}%
  \BibitemOpen
  \href@noop {} {}\bibinfo {note} {See Supplementary Materials, and references
  therein.}\BibitemShut {Stop}%
\bibitem [{\citenamefont {Kantorovich}(1960)}]{kantorovich1960mathematical}%
  \BibitemOpen
  \bibfield  {author} {\bibinfo {author} {\bibfnamefont {L.~V.}\ \bibnamefont
  {Kantorovich}},\ }\bibfield  {title} {\bibinfo {title} {Mathematical methods
  of organizing and planning production},\ }\href
  {https://pubsonline.informs.org/doi/abs/10.1287/mnsc.6.4.366} {\bibfield
  {journal} {\bibinfo  {journal} {Management science}\ }\textbf {\bibinfo
  {volume} {6}},\ \bibinfo {pages} {366} (\bibinfo {year} {1960})}\BibitemShut
  {NoStop}%
\bibitem [{\citenamefont {Vaserstein}(1969)}]{vaserstein1969markov}%
  \BibitemOpen
  \bibfield  {author} {\bibinfo {author} {\bibfnamefont {L.~N.}\ \bibnamefont
  {Vaserstein}},\ }\bibfield  {title} {\bibinfo {title} {Markov processes over
  denumerable products of spaces, describing large systems of automata},\
  }\href
  {https://www.mathnet.ru/php/archive.phtml?wshow=paper&jrnid=ppi&paperid=1811&option_lang=eng}
  {\bibfield  {journal} {\bibinfo  {journal} {Problemy Peredachi Informatsii}\
  }\textbf {\bibinfo {volume} {5}},\ \bibinfo {pages} {64} (\bibinfo {year}
  {1969})}\BibitemShut {NoStop}%
\bibitem [{\citenamefont {Larson}\ \emph {et~al.}(2025)\citenamefont {Larson},
  \citenamefont {Mondaini},\ and\ \citenamefont {Scalettar}}]{zenodo}%
  \BibitemOpen
  \bibfield  {author} {\bibinfo {author} {\bibfnamefont {R.}~\bibnamefont
  {Larson}}, \bibinfo {author} {\bibfnamefont {R.}~\bibnamefont {Mondaini}},\
  and\ \bibinfo {author} {\bibfnamefont {R.}~\bibnamefont {Scalettar}},\
  }\bibfield  {title} {\bibinfo {title} {Dataset for `{Sign-Resolved Statistics
  and the Origin of Bias in Quantum Monte Carlo}'},\ }\href
  {https://doi.org/10.5281/zenodo.17782714} {10.5281/zenodo.17782714} (\bibinfo
  {year} {2025})\BibitemShut {NoStop}%
\bibitem [{\citenamefont {Huang}\ \emph {et~al.}(2018)\citenamefont {Huang},
  \citenamefont {Mendl}, \citenamefont {Jiang}, \citenamefont {Moritz},\ and\
  \citenamefont {Devereaux}}]{huang2018stripe}%
  \BibitemOpen
  \bibfield  {author} {\bibinfo {author} {\bibfnamefont {E.~W.}\ \bibnamefont
  {Huang}}, \bibinfo {author} {\bibfnamefont {C.~B.}\ \bibnamefont {Mendl}},
  \bibinfo {author} {\bibfnamefont {H.-C.}\ \bibnamefont {Jiang}}, \bibinfo
  {author} {\bibfnamefont {B.}~\bibnamefont {Moritz}},\ and\ \bibinfo {author}
  {\bibfnamefont {T.~P.}\ \bibnamefont {Devereaux}},\ }\bibfield  {title}
  {\bibinfo {title} {Stripe order from the perspective of the {H}ubbard
  model},\ }\href {https://www.nature.com/articles/s41535-018-0097-0.pdf}
  {\bibfield  {journal} {\bibinfo  {journal} {npj Quantum Materials}\ }\textbf
  {\bibinfo {volume} {3}},\ \bibinfo {pages} {22} (\bibinfo {year}
  {2018})}\BibitemShut {NoStop}%
\bibitem [{\citenamefont {Xu}\ \emph {et~al.}(2024)\citenamefont {Xu},
  \citenamefont {Chung}, \citenamefont {Qin}, \citenamefont {Schollwöck},
  \citenamefont {White},\ and\ \citenamefont {Zhang}}]{Xu2024}%
  \BibitemOpen
  \bibfield  {author} {\bibinfo {author} {\bibfnamefont {H.}~\bibnamefont
  {Xu}}, \bibinfo {author} {\bibfnamefont {C.-M.}\ \bibnamefont {Chung}},
  \bibinfo {author} {\bibfnamefont {M.}~\bibnamefont {Qin}}, \bibinfo {author}
  {\bibfnamefont {U.}~\bibnamefont {Schollwöck}}, \bibinfo {author}
  {\bibfnamefont {S.~R.}\ \bibnamefont {White}},\ and\ \bibinfo {author}
  {\bibfnamefont {S.}~\bibnamefont {Zhang}},\ }\bibfield  {title} {\bibinfo
  {title} {Coexistence of superconductivity with partially filled stripes in
  the {H}ubbard model},\ }\href {https://doi.org/10.1126/science.adh7691}
  {\bibfield  {journal} {\bibinfo  {journal} {Science}\ }\textbf {\bibinfo
  {volume} {384}},\ \bibinfo {pages} {eadh7691} (\bibinfo {year}
  {2024})}\BibitemShut {NoStop}%
\bibitem [{\citenamefont {Bhattacharyya}(1946)}]{bhattacharyya1946measure}%
  \BibitemOpen
  \bibfield  {author} {\bibinfo {author} {\bibfnamefont {A.}~\bibnamefont
  {Bhattacharyya}},\ }\bibfield  {title} {\bibinfo {title} {On a measure of
  divergence between two multinomial populations},\ }\href
  {https://www.jstor.org/stable/25047882} {\bibfield  {journal} {\bibinfo
  {journal} {Sankhy{\=a}: the {I}ndian journal of statistics}\ ,\ \bibinfo
  {pages} {401}} (\bibinfo {year} {1946})}\BibitemShut {NoStop}%
\end{thebibliography}%

\clearpage

% Number figures, tables, equations as S1, S2, ...
\renewcommand{\thefigure}{S\arabic{figure}}
\renewcommand{\thetable}{S\arabic{table}}
\renewcommand{\theequation}{S\arabic{equation}}
\setcounter{figure}{0}
\setcounter{table}{0}
\setcounter{equation}{0}

% Number sections as S1, S2, S3, ...
\renewcommand{\thesection}{S\arabic{section}}
\setcounter{section}{0}

\onecolumngrid
\section*{Supplemental Material\\[3pt]
for ``Sign-Resolved Statistics and the Origin of Bias in Quantum Monte Carlo''}

These supplementary materials present additional results, including definitions of certain physical observables, distributions of equal-time pair correlations, extended data for metrics associated with histogram deviations, including a comparison to the Bhattacharyya distance, an analysis of finite-size and finite imaginary-time discretization effects in the main results, and a derivation of Eq.~\eqref{eq:sign_no_sign_relation} of the main text.

\twocolumngrid
% \vskip 0.5in

\section{S1. Spin and Pairing Correlation Functions}

In the main text, we use various physical quantities to understand how different signs of the weight distributions affect them; let us define them appropriately. Long-range order in the repulsive Hubbard model is determined by the real-space spin
\begin{align}
c_j({\bf r}) = \langle \hat S^{-}_{j+{\bf r}} \hat S^{+}_{j} \rangle 
 = \langle \, \hat c^{\dagger}_{j+{\bf r}\downarrow} 
 \, \hat c^{\phantom{\dagger}}_{j+{\bf r}\uparrow} \, 
 \, \hat c^{\dagger}_{j\uparrow} \,
 \hat c^{\phantom{\dagger}}_{j\downarrow} \, \rangle \ ,
\end{align}
and pairing correlation functions
\begin{align}
p_j^{(\alpha)}({\bf r}) &= \langle \, \hat \Delta^{(\alpha) \phantom{\dagger}}_{j+{\bf r}} 
\, \hat \Delta^{(\alpha) \dagger}_{j} \rangle
\nonumber \\
\hat  \Delta^{(s)}_{j} &= \hat c^{\phantom{\dagger}}_{j\downarrow} \, 
 \hat c^{\phantom{\dagger}}_{j\uparrow} \, 
\nonumber \\
\hat \Delta^{(s^*)}_{j} &= 
\frac{1}{2} ( \, \hat c^{\phantom{\dagger}}_{j+ \hat x\downarrow} 
 + \hat c^{\phantom{\dagger}}_{j-\hat x\downarrow} 
 + \hat c^{\phantom{\dagger}}_{j+\hat y\downarrow} 
 + \hat c^{\phantom{\dagger}}_{j-\hat y\downarrow}  \, )
 \, \hat c^{\phantom{\dagger}}_{j\uparrow} 
\nonumber \\
 \hat \Delta^{(d)}_{j} &= 
\frac{1}{2} ( \, \hat c^{\phantom{\dagger}}_{j+\hat x\downarrow} 
 + \hat c^{\phantom{\dagger}}_{j-\hat x\downarrow} 
 - \hat c^{\phantom{\dagger}}_{j+\hat y\downarrow} 
 - \hat c^{\phantom{\dagger}}_{j-\hat y\downarrow}  \, )
 \, \hat c^{\phantom{\dagger}}_{j\uparrow} \ ,
\end{align} 
where $c_j(r)$ samples the $x$ and $y$ spin correlations. We also measure the $z$ direction spin correlator, $\langle \hat S^{z}_{j+{\bf r}} \hat S^{z}_{j}\rangle$, which, within statistical error bars, is equal to its $x/y$ counterpart owing to the SU(2) symmetry. For the pairing correlations, $\alpha$ labels three possible symmetries investigated, $s,s^*,d$, corresponding to isotropic $s$-wave pairing, extended $s$-wave and $d$-wave pairing symmetries~\cite{white89a}.

\begin{figure}[t]
\includegraphics[width=1.0\columnwidth]{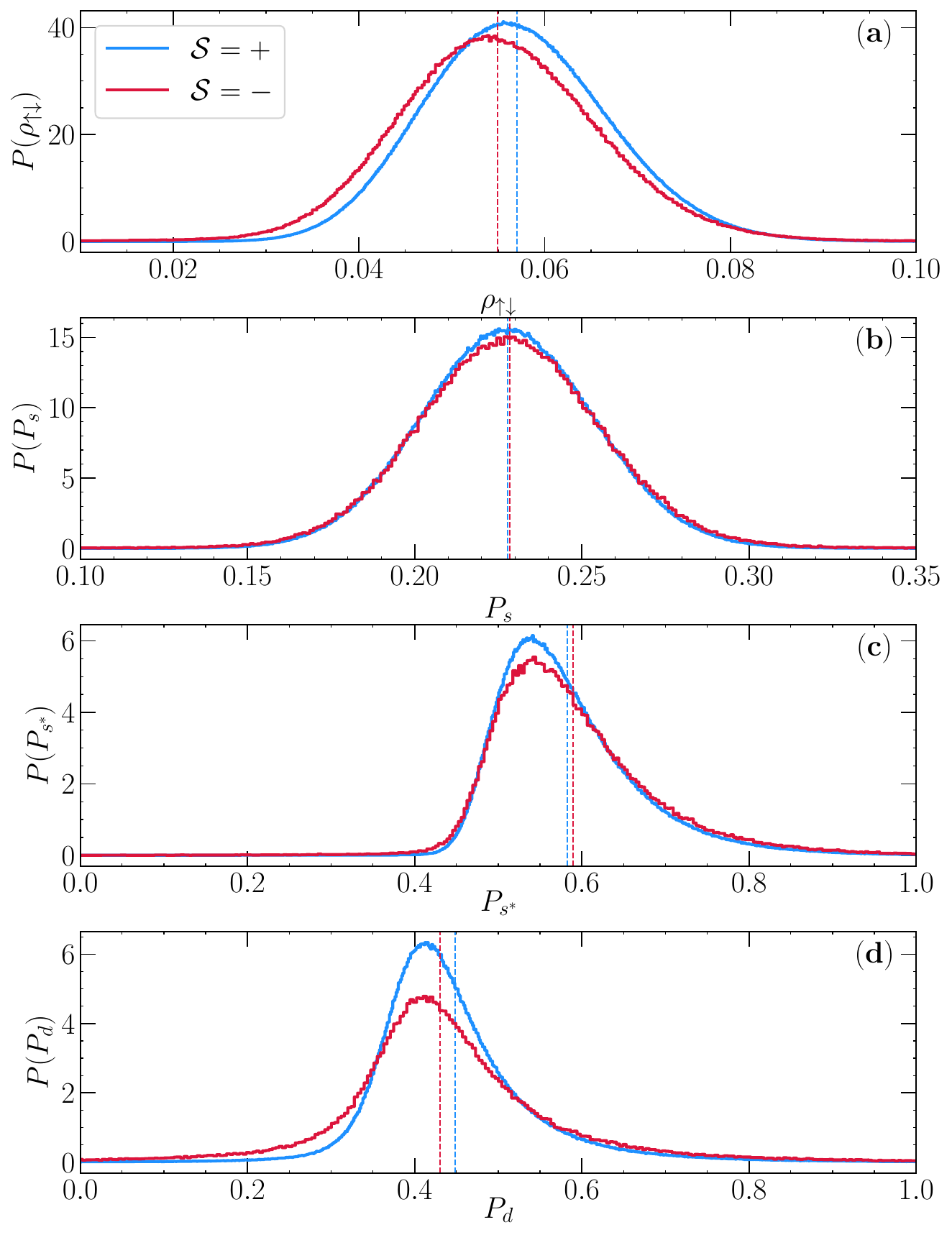}
    \caption{Histograms of (a) the double occupancy $\rho_{\uparrow\downarrow}=\langle \hat n_{\uparrow} \hat n_{\downarrow} \rangle$, and (b-d) the {\it equal time} pair structure factors $P_s, P_{s^*}, P_d$.  Parameters are the same as in Fig.~\ref{fig:fig_1_mod} of the main text: $8\times 8$ spatial lattice with $U/t = 6$, $T/t=1/3$ and $\mu/t=-1.4$; this leads to a mean density  $\rho \simeq 0.88$ with $\langle {\cal S} \rangle \simeq 0.83$. The imaginary-time discretization is set at $t\Delta\tau = 0.05$.}
    \label{fig:figS1}
\end{figure}
 
The magnetic and pairing structure factors are the Fourier transforms of the above correlations,
\begin{align}
S({\bf q}) &= \sum_{j,{\bf r}} e^{i {\bf q} \cdot {\bf r}} c_j({\bf r})
\nonumber \\
P_{\alpha}({\bf q}) &= \sum_{j,{\bf r}} e^{i {\bf q} \cdot {\bf r}} p_j^{(\alpha)}({\bf r})\ .
\end{align}
We will focus on the antiferromagnetic structure factor $S_{\rm \scriptscriptstyle AF} =  S({\bf q}=(\pi,\pi))$ and the uniform pairing structure factors $P_{\alpha}({\bf q}=(0,0)) $. Finally, the associated susceptibilities $\chi_\alpha$ generalize the above expressions to include a time separation $\tau$ of the operators whose correlation is being measured, in addition to the spatial separation $r$, and then integrate over $\tau$:
\begin{align}
\chi_\alpha &= \frac{1}{L^2} 
\sum_{j,{\bf r}} \int_{0}^{\beta} d\tau \;
\Big\langle 
\, \hat \Delta^{(\alpha)}_{j+{\bf r}}(\tau) \,
\hat \Delta^{(\alpha)\dagger}_{j}(0) 
\Big\rangle ,
\end{align}
where $\hat \Delta^{(\alpha)}_{j}(\tau)=e^{\tau \hat{\cal H}} \hat \Delta^{(\alpha)}_{j} e^{-\tau \hat{\cal H}}$
is the Heisenberg-picture pair annihilation operator.  
This quantity measures the integrated pairing correlations in both space and imaginary time and 
captures the low-energy, long-wavelength pairing fluctuations appropriate for assessing 
superconducting tendencies.

 \begin{figure}[t]
 \includegraphics[width=1\columnwidth]{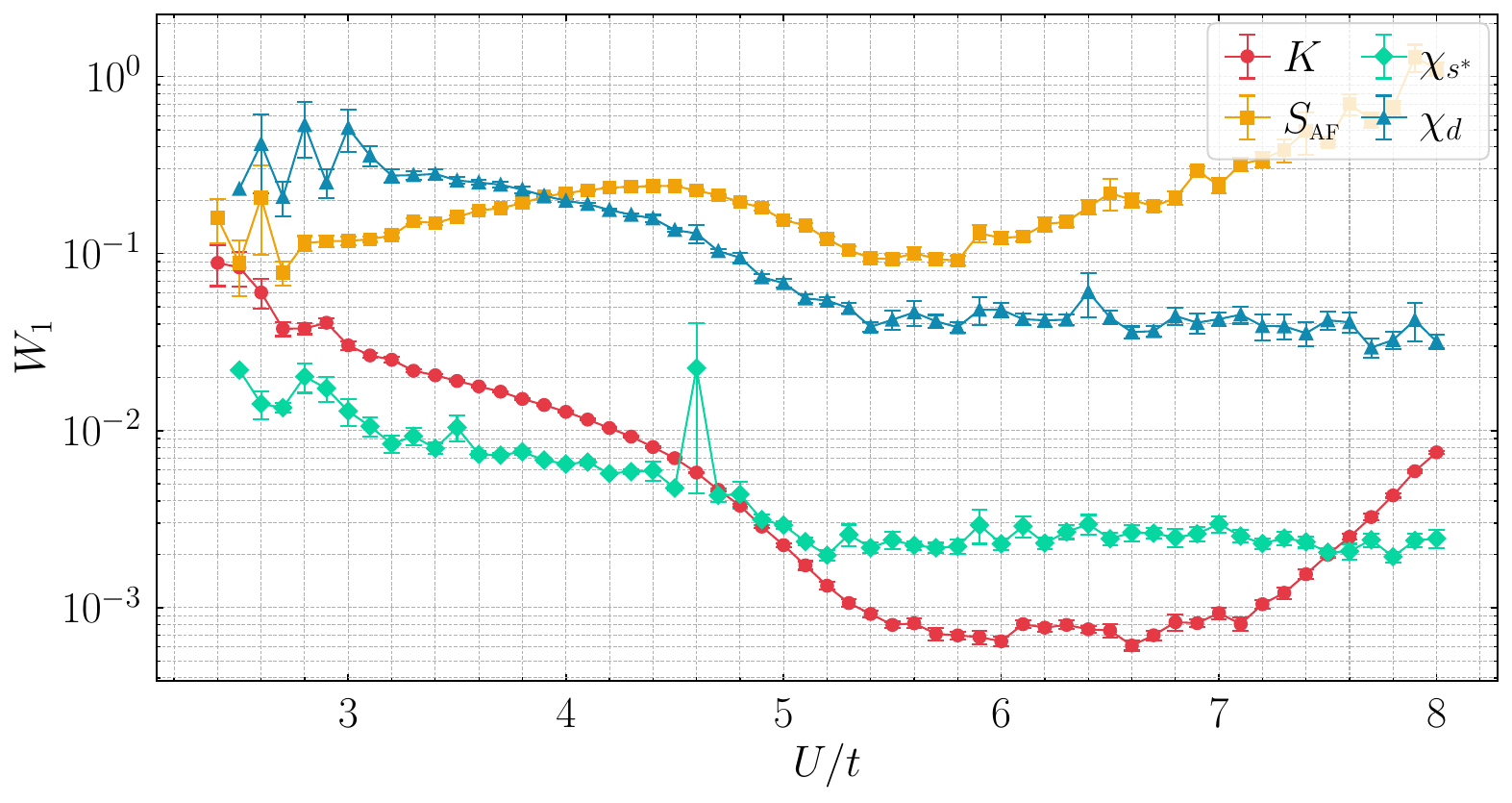}
 \caption{
 Wasserstein distance between the sign resolved distributions $P_+$ and $P_-$ as a function of the interaction strength $U$ at a fixed temperature $T/t = 0.2$ for an $8\times 8$ spatial lattice at chemical potential $\mu/t=-1.4$. Results are averaged over 24 independent Markov chains, and the error bars are the standard error of the mean. Here, the results are not normalized by the standard deviation, unlike in the main text, to emphasize the larger dissimilarity of $P_\pm$ distributions for the observables $S_{\scriptscriptstyle \rm AF}$ and $\chi_d$, in direct correspondence to what is originally seen in Fig.~\ref{fig:fig_1_mod}
of the main text.}
 \label{fig:figS2}
 \end{figure}

\section{S2.  Additional Histograms}

In the main text, Fig.~\ref{fig:fig_1_mod} showed histograms for the kinetic energy, (equal time) antiferromagnetic structure factor, and $s^*$- and $d$-wave pairing susceptibilities $\chi_\alpha$. We chose to show the imaginary-time integrated quantities in the case of pairing because signals of such off-diagonal order are generally less strong than diagonal order, and hence $\chi_{\alpha}$ provides a more sensitive probe of superconductivity. For completeness, Fig.~\ref{fig:figS1} panels (b,c,d) also exhibit histograms for the equal time pair correlations. As in the case of $\chi_\alpha$, the largest difference between the $P_+$ and $P_-$ distributions occurs in the $d$-wave channel, although overall the equal-time quantities exhibit smaller sign-sector dissimilarities than their imaginary-time-integrated counterparts. Finally, Fig.~\ref{fig:figS1}(a) shows the histogram of the double occupancy $\rho_{\uparrow\downarrow}=\langle \hat n_\uparrow \hat n_\downarrow\rangle$, which directly contributes to the potential-energy term $\propto U\rho_{\uparrow\downarrow}$.

\section{S3.  Wasserstein Distance Formal Definition}
In the main text, we employ the Wasserstein distance, a metric to identify how different the distributions $ P_{+} ({\cal O})$ and $P_-({\cal O})$ are. Let us now formally define it. Given two probability distributions, $\pi_{j}(x),\, j=1,2$ the ${\cal WD}$ is given by ${\cal W}_1 \equiv \int | \Pi_{1}(x) - \Pi_{2}(x)| \, dx$, where $\Pi_{j}(x)= \int_{-\infty}^{x} \pi_{j}(x^{\prime}) \, dx^{\prime}$ are the cumulative probability distributions of $\pi_{j}(x)$. The virtue of the WD is that given a list of values $\{x_{i}\}$ for $i=1,\ldots, N$ the cumulative distribution is naturally defined as a set of steps of size $1/N$ as $x$ passes through each $x_{i}$, and the calculation of ${\cal W}_{1}$ can be performed without introducing any {\it ad hoc} bin size. In our case, the lists are of the measurements of an observable ${\cal O}$ in the ${\cal S}=\pm1$ sectors.

To illustrate how the histogram dissimilarity varies across interaction strengths, Fig.~\ref{fig:figS2} shows $W_1/\sigma_{\rm tot}$ as a function of $U/t$ at fixed $T/t = 0.2$. Because changing $U$ at fixed $\mu$ and $T$ simultaneously modifies the density, the shapes of the sign-resolved distributions, and the average sign, the resulting trends in $W_1$ cannot be attributed to a single monotonic physical mechanism. Instead, the figure shows that the dissimilarity between sign sectors depends nontrivially on $U$, reflecting the intricate interplay among interaction strength, filling, and the structure of the fermion determinants in DQMC. Notably, the regime where the distributions are most similar corresponds to the value of interactions ($U/t=6$) used in the main text, which focuses on the temperature dependence.

 \begin{figure}[t]
 \includegraphics[width=1\columnwidth]{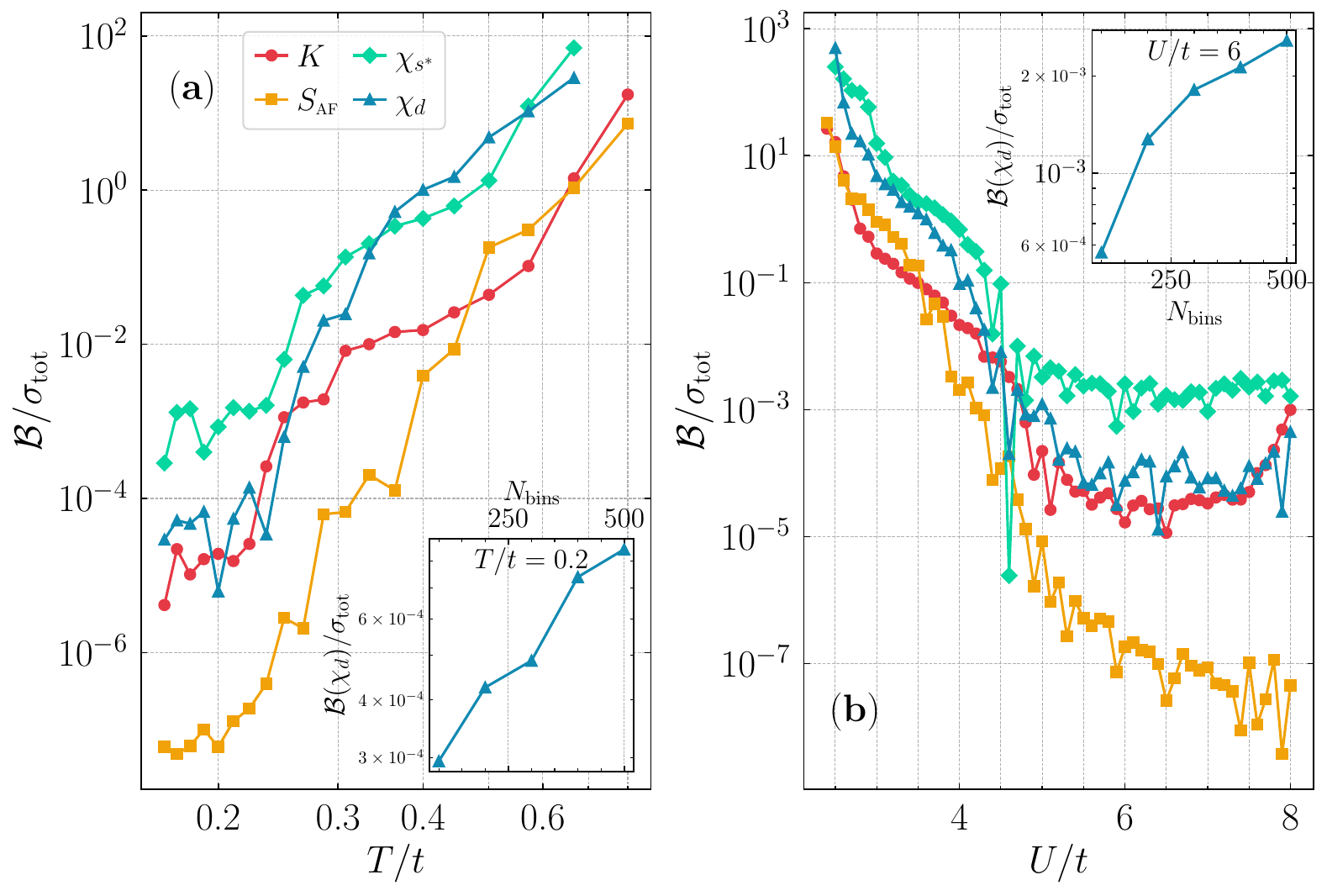}
 \caption{
 (a) Temperature dependence of the Bhattacharyya distance ${\cal B}$ for the same observables in Fig.~\ref{fig:fig_3} of the main text, using $U/t=6$. (b) Interaction dependence of the Bhattacharyya distance ${\cal B}$ at $T/t=0/2$. Here, the chemical potential is set at $\mu/t=-1.4$. In all cases $\cal B$ is normalized by the standard deviation of the combined ${\cal S}=\pm1$ distribution, and the imaginary time discretization is set at $t\Delta\tau=0.05$. Insets in (a) and (b) show the dependence of ${\cal B}$ for $d$-wave susceptibility $\chi_d$ on the number of bins used in the histogram, at representative temperature and interaction strength, respectively.}
 \label{fig:figS3}
 \end{figure}

\section{S4. Bhattacharyya Distance}

An alternative measure of dissimilarity between probability distributions is the Bhattacharyya distance~\cite{bhattacharyya1946measure}. Although the Wasserstein distance used in the main text has the advantage of being independent of binning choices, it is useful to confirm that our conclusions do not depend on the particular metric employed. As we show here, the Bhattacharyya distance yields qualitatively identical trends.

For two normalized histograms $\pi_1(i)$ and $\pi_2(i)$, the Bhattacharyya distance is defined as
\begin{align}
{\cal B} \equiv -\ln \left( \sum_i \sqrt{\pi_1(i) \ \pi_2(i)} \right)\ .
\end{align}
This definition satisfies two desirable limits: ${\cal B}=0$ when $\pi_1(i)=\pi_2(i)$ for all $i$, and ${\cal B}\to\infty$ when the supports of $\pi_1$ and $\pi_2$ do not overlap, i.e., when $\pi_1(i)$ is nonzero only where $\pi_2(i)=0$ and \emph{vice versa}.

Figure~\ref{fig:figS3} shows ${\cal B}$ for the same parameters used in Fig.~\ref{fig:fig_3} of the main text and Fig.~\ref{fig:figS2}: fixed chemical potential $\mu/t=-1.4$ for an $8\times8$ lattice. The left panel displays the temperature dependence, which decreases as $T$ is lowered, while the right panel shows the interaction-strength dependence. In both cases, the qualitative behavior mirrors that of the normalized Wasserstein distance ${\cal W}_1/\sigma_{\rm tot}$. This agreement demonstrates that our conclusions concerning the sign-resolved dissimilarity of observables are robust and do not depend on the specific choice of histogram-distance metric.

\section{S5. Proof of Eq.~(5)}

Equation~\eqref{eq:sign_no_sign_relation} of the main text states that the bias incurred by ignoring the fermion sign in a Monte Carlo measurement is
\begin{equation}
\langle {\cal O}\rangle_{\cal W} - \langle {\cal O}\rangle_{|{\cal W}|}
    = \frac{\Delta\mu\,(1-\langle {\cal S}\rangle^{2})}{2\langle {\cal S}\rangle},
\label{eq:Sbiasrelation}
\end{equation}
where $\Delta\mu=\mu_{+}-\mu_{-}$ is the difference of the means of the sign-resolved distributions $P_{\pm}({\cal O})$, and $\langle {\cal S}\rangle=P_{+}-P_{-}$ is the average sign under the $|{\cal W}|$ sampling measure. Here we provide a compact derivation.

Under sampling with the non-negative weight $|{\cal W}|$, let $P_{\pm}$ denote the probabilities of encountering configurations with ${\cal S}=\pm 1$, and let $\mu_{\pm}=\langle {\cal O}\rangle_{\pm}$ be the corresponding conditional means. By definition,
\begin{align}
\langle {\cal O}\rangle_{|{\cal W}|}
    &= P_{+}\,\mu_{+} + P_{-}\,\mu_{-}\ \ , \label{eq:Sabs}\\[4pt]
\langle {\cal O}\rangle_{\cal W}
    &= \frac{P_{+}\,\mu_{+} - P_{-}\,\mu_{-}}{P_{+}-P_{-}}\ \ .
\label{eq:Ssign}
\end{align}

Subtracting Eqs.~(\ref{eq:Sabs})–(\ref{eq:Ssign}) yields
 \begin{align}
& \langle {\cal O}\rangle_{\cal W}-\langle{{\cal O}}\rangle_{|{\cal W}|}  
\nonumber \\
&= \frac{P_+\mu_+-P_-\mu_--(P_+-P_-)(P_+\mu_++P_-\mu_-)}{P_+-P_-} 
\nonumber \\
&= \frac{\mu_+(P_++P_+P_--P_+^2)+\mu_-(-P_--P_+P_-+P_-^2)}{P_+-P_-} 
\nonumber \\
&= \frac{\mu_+P_+(1-P_++P_-)-\mu_-P_-(1+P_+-P_-)}{P_+-P_-} 
\nonumber \\
&= \frac{\mu_+P_+(2P_-)-\mu_-P_-(2P_+)}{P_+-P_-} 
\nonumber \\
&= \frac{(2P_+P_-)\Delta\mu}{P_+-P_-}
\,\,,
\label{eq:Sintermediate}
\end{align}

Using $P_{+}+P_{-}=1$ and $\langle {\cal S}\rangle=P_{+}-P_{-}$, one finds
\[
P_{+} = \frac{1+\langle {\cal S}\rangle}{2},
\qquad
P_{-} = \frac{1-\langle {\cal S}\rangle}{2},
\qquad
P_{+}P_{-}=\frac{1-\langle {\cal S}\rangle^{2}}{4}.
\]
Substituting this into Eq.~(\ref{eq:Sintermediate}) gives
\begin{equation}
\langle {\cal O}\rangle_{\cal W} - \langle {\cal O}\rangle_{|{\cal W}|}
    = \frac{\Delta\mu\,(1-\langle {\cal S}\rangle^{2})}{2\,\langle {\cal S}\rangle}\ ,
\end{equation}
which completes the proof of Eq.\,\eqref{eq:sign_no_sign_relation}.

\section{S6.  Resolution with ${\cal S}_{\uparrow}$ and ${\cal S}_{\downarrow}$ }

\begin{figure}[t]
\includegraphics[width=1\columnwidth]{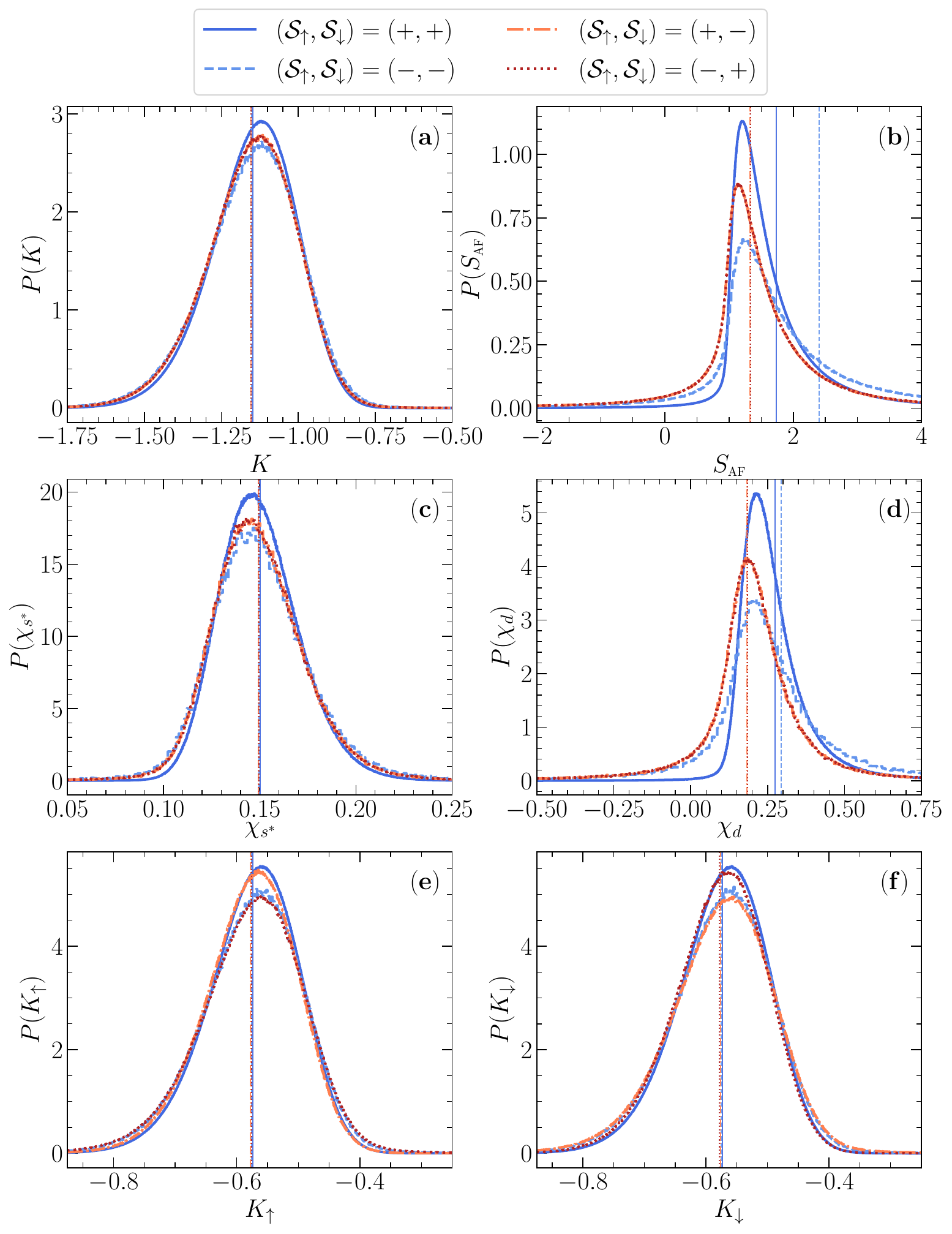}
\caption{Histograms resolved by the spin-resolved signs $({\cal S}_{\uparrow},{\cal S}_{\downarrow})$ for (a) total kinetic energy $K$, (b) antiferromagnetic structure factor $S_{\scriptscriptstyle\rm AF}$, (c) extended-$s$ pairing susceptibility $\chi_{s^*}$, (d) $d$-wave pairing susceptibility $\chi_d$, and spin-resolved kinetic energies (e) $K_{\uparrow}$ and (f) $K_{\downarrow}$.}
\label{fig:figS4}
\end{figure}

In DQMC, the weight is the product of two determinants, one resulting from the fermion trace of the up-spin fermions and one from the down-spin fermions. A sign is associated with each. While the main text separated measurements only by the {\it total} sign ${\cal S}={\cal S}_{\uparrow}{\cal S}_{\downarrow}=(+,-)$, a finer decomposition is possible by examining the four spin-resolved sectors $({\cal S}_{\uparrow},{\cal S}_{\downarrow})\in\{(+,+),(+,-),(-,+),(-,-)\}$. Figure~\ref{fig:figS4} shows this resolution for several observables: the total kinetic energy $K$, the antiferromagnetic structure factor $S_{\scriptscriptstyle\rm AF}$, the extended-$s$ and $d$-wave pairing susceptibilities $\chi_{s^*}$ and $\chi_d$, and the spin-resolved kinetic energies $K_{\uparrow}$ and $K_{\downarrow}$.

For observables that are spin symmetric (such as density, double occupancy, total kinetic energy, magnetic correlations, and the pairing susceptibilities), the histograms for $(+,-)$ and $(-,+)$ must be identical. This SU(2) symmetry is clearly borne out in panels (a)--(d) of Fig.~\ref{fig:figS4}. In contrast, the two sectors with {\it total positive sign}, $({\cal S}_{\uparrow},{\cal S}_{\downarrow})=(+,+)$ and $(-,-)$, need not produce identical histograms, since the up- and down-spin determinants fluctuate independently. Indeed, the histograms differ noticeably for $S_{\scriptscriptstyle\rm AF}$, $\chi_{s^*}$, $\chi_d$, and $K$, with the $({\cal S}_{\uparrow},{\cal S}_{\downarrow})=(+,+)$ sector generally exhibiting a more sharply peaked distribution, especially in the $d$-wave channel.

Panels (e) and (f) highlight this further by showing $K_{\uparrow}$ and $K_{\downarrow}$ separately in the four $({\cal S}_{\uparrow},{\cal S}_{\downarrow})$ sectors. For the {\it total} kinetic energy $K=K_{\uparrow}+K_{\downarrow}$, SU(2) symmetry and exchange of spin labels enforce identical histograms in the $(+,-)$ and $(-,+)$ sectors, as seen in panel (a). However, for the spin-resolved quantities $K_{\uparrow}$ and $K_{\downarrow}$, this constraint no longer applies. Conditioning on $({\cal S}_{\uparrow},{\cal S}_{\downarrow})=(+,-)$ or $(-,+)$ selects different subsets of Hubbard-Stratonovich configurations, so the distributions of $K_{\uparrow}$ in these two sectors need not coincide (and likewise for $K_{\downarrow}$). The differences visible in panels (e) and (f), therefore, reflect nontrivial correlations between the spin-resolved determinants and spin-resolved kinetic energies, even though the distribution of the {\it sum} $K$ is the same in the $(+,-)$ and $(-,+)$ sectors.

Overall, the spin-resolved histograms obey all required symmetries, and the residual differences between curves are consistent with statistical uncertainties based on the sampling time and the smoothness expected from the underlying distributions.

\subsection{S7. Finite-Size and Trotter Dependence}
\begin{figure*}[t!]
\includegraphics[width=0.72\textwidth]{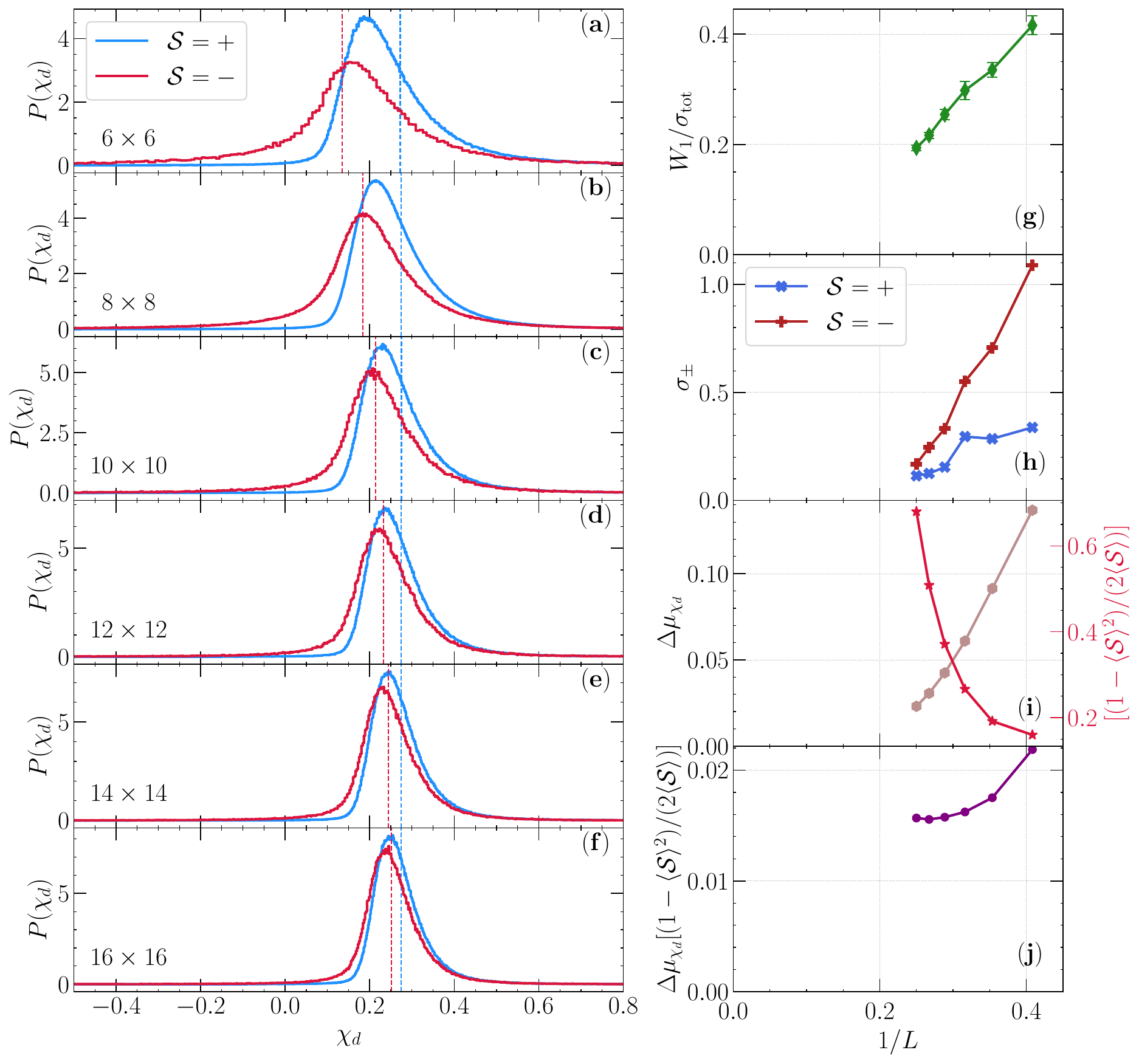}
\caption{
(a--f) Finite-size effects on the histograms of the $d$-wave pairing susceptibility $\chi_d$ for $U/t=6$, $\beta=3/t$, and $\rho \simeq 0.875$. System sizes $6\times6$, $8\times8$, $10\times10$, $12\times12$, $14\times14$, and $16\times 16$ have $\langle {\cal S} \rangle$ $\approx$ 0.85, 0.83, 0.77, 0.70, 0.61, and 0.53, respectively. (g) Normalized Wasserstein distance $W_1/\sigma_{\rm tot}$ between the $P_{\pm}(\chi_d)$ distributions as a function of $1/L$, showing that they exhibit a systematic decrease with increasing $L$, suggesting a vanishing difference in approaching the thermodynamic limit. (h) Standard deviations $\sigma_{\pm}$ of the two sign-resolved distributions versus $1/L$, highlighting that the ${\cal S}=-$ sector exhibits a larger intrinsic width, and that they both shrink with $L$. (i) Difference of the mean susceptibilities, $\Delta\mu_{\chi_d}=\mu_+ - \mu_-$, as a function of $1/L$. The red curve (right vertical axis) shows the amplification factor $(1-\langle{\cal S}\rangle^2)/(2\langle{\cal S}\rangle)$. (j) The product of $\Delta\mu_{\chi_d}$ and the amplification factor, which saturates for large $L$, indicates that the bias between sign-resolved and sign-ignored estimates of the $d$-wave pairing susceptibility remains finite in the thermodynamic limit.} 
\label{fig:figS5}
\end{figure*}
In this final subsection of the SM, we analyze the effects of the finite spatial lattice size and the Trotter discretization on our results.

Figure~\ref{fig:figS5}(a--f) compares sign-resolved histograms of $\chi_d$ for $6 \times 6$ to $16 \times 16$ lattices at $U/t=6$, $T/t=1/3$, and density $\rho \simeq 0.875$ (obtained by tuning $\mu$ for each $L$). At first glance, the $P_\pm(\chi_d)$ distributions appear to become more similar as $L^2$ increases. However, one must account for the fact that the histograms become trivially narrower as the system size increases, due to averaging over more sites. This is quantified in Fig.~\ref{fig:figS5}(h), which shows the standard deviations $\sigma_\pm$ of $P_\pm$: both widths decrease with $1/L$, and the ${\cal S}=-$ sector consistently exhibits a larger intrinsic spread.

A more meaningful comparison is therefore provided by dimensionless measures. One such quantity is the normalized Wasserstein distance $W_1/\sigma_{\rm tot}$, shown in Fig.~\ref{fig:figS5}(g). Since the distributions become closer as $L$ grows, this is reflected in $W_1/\sigma_{\rm tot}$, which does exhibit a clear trend toward zero within the accessible range of system sizes, suggesting that a vanishing dissimilarity between $P_+$ and $P_-$ distributions. A complementary measure is the difference in the means, normalized by the combined width,
\begin{equation}
  \frac{\Delta \mu_{\chi_d}}{\sigma_{\rm tot}}
  \equiv \frac{\langle \chi_d \rangle_+ - \langle \chi_d \rangle_-}{\sigma_{\rm tot}} \, ,
\end{equation}
whose values are listed in Table~S1. While this quantity decreases from $6\times 6$ to $10\times 10$, it does not continue to diminish systematically; for the largest three sizes, it fluctuates around $\sim 0.2$, consistent with a finite asymptotic value rather than a vanishing one.

\begin{table}[htpb]
\centering
\label{tab:my_table}
\begin{tabular}{c c}
\hline\hline
$N$ & \hskip0.15in $\Delta \mu_{\chi_{d}}/\sigma_{\rm tot}$ \\
\hline
$6\times 6$   & \hskip0.15in 0.31 \\
$8\times 8$   & \hskip0.15in 0.27 \\
$10\times 10$ & \hskip0.15in 0.18 \\
$12\times 12$ & \hskip0.15in 0.22 \\
$14\times 14$ & \hskip0.15in 0.20 \\
$16\times 16$ & \hskip0.15in 0.18 \\
\hline\hline
\end{tabular}
\caption{
The normalized difference in means $\Delta \mu_{\chi_{d}} \equiv  \big(\langle \chi_{d} \rangle_+ - \langle \chi_{d} \rangle_-\big) / \sigma_{\rm tot}$ of the $d$-wave pair susceptibility distributions, normalized to the width of the combined (total) distribution, as a function of system size $N$.
}
\end{table}

This finite-size analysis also allows us to examine how the measurement bias, as encoded in Eq.~\eqref{eq:sign_no_sign_relation} of the main text, scales with $L$. In particular, Fig.~\ref{fig:figS5}(i) shows the size dependence of the unnormalized mean difference $\Delta\mu_{\chi_d}=\mu_+ - \mu_-$ (left axis) and of the amplification factor $(1-\langle{\cal S}\rangle^2)/(2\langle{\cal S}\rangle)$ (right axis). As expected, $\Delta\mu_{\chi_d}$ decreases with increasing $L$, while the amplification factor grows due to the corresponding decay of $\langle{\cal S}\rangle$. Their product,
\[
  \Delta\mu_{\chi_d} \times \frac{1-\langle{\cal S}\rangle^2}{2\langle{\cal S}\rangle} \, ,
\]
which directly estimates the bias $\langle\chi_d\rangle_{\cal W} - \langle\chi_d\rangle_{|{\cal W}|}$ [see Eq.~\eqref{eq:sign_no_sign_relation} of the main text], is plotted in Fig.~\ref{fig:figS5}(j) as a function of $1/L$. Within our numerical resolution, this quantity approaches a size-independent plateau for the largest lattices, indicating that the measurement bias remains finite in the thermodynamic limit rather than being a finite-size artifact.

\begin{figure}[t]
\includegraphics[width=1\columnwidth]
{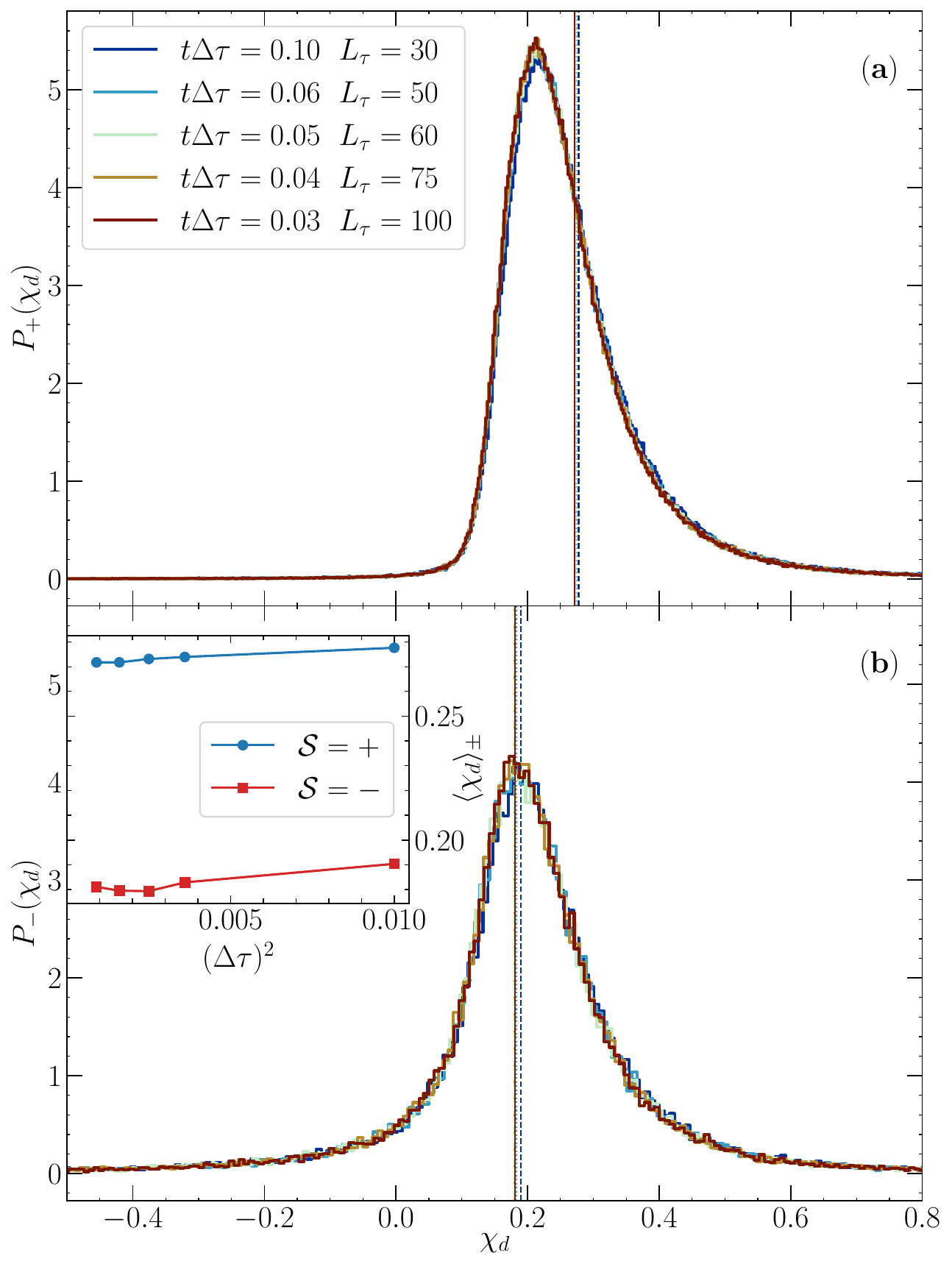}
\caption{Sign-resolved histograms of the $d$-wave pairing susceptibility $\chi_d$ for different imaginary-time discretizations at $\beta t=3$ for an $8\times 8$ lattice at $U/t=6$. (a) [(b)] probability distribution $P_+(\chi_d)$ [$P_-(\chi_d)$] obtained from configurations with positive [negative] Monte Carlo weights for several values of $t\Delta\tau$; vertical lines indicate the corresponding mean values $\mu_\pm$. The systematic shift between the ${\cal S}=+1$ and ${\cal S}=-1$ sectors persists for all discretizations. The inset displays the mean values $\langle\chi_d\rangle_\pm$ as a function of $(\Delta\tau)^2$, showing the expected linear approach to the $\Delta\tau\to0$ limit for both sign sectors. These results demonstrate that the sign-resolved means are well behaved under Trotter extrapolation and that the bias $\Delta\mu$ remains finite even as $\Delta\tau\to0$.}
\label{fig:figS6}
\end{figure}

Finally, to assess possible effects of the Trotter discretization, Fig.~\ref{fig:figS6} compares sign-resolved histograms of $\chi_d$ for several combinations of imaginary-time slices $L_\tau$ and discretizations $\Delta\tau$ such that their product, the inverse temperature $\beta=L_\tau \Delta\tau$, is fixed at $\beta t=3$ for an $8\times 8$ lattice at $U/t=6$. The $P_\pm(\chi_d)$ histograms are statistically indistinguishable across the different $t\Delta\tau$ values, and the inset shows that the sign-resolved means $\langle\chi_d\rangle_\pm$ vary linearly with $(\Delta\tau)^2$ and extrapolate smoothly to distinct $\Delta\tau\to 0$ limits. These results demonstrate that the sign-resolved means are well behaved under Trotter extrapolation, and that the bias $\Delta\mu_{\chi_d}$ remains finite even in the continuum-time limit within our statistical accuracy.

\end{document}